# Tuning the Magnetic Properties of M-type Hexaferrites

Sami .H. Mahmood[1,a], Ibrahim Bsoul[2,b]

[1]Physics Department, The University of Jordan, Amman 11942, Jordan
[2]Physics Department, Al al-Bayt University, Mafraq 13040, Jordan

[a]s.mahmood@ju.edu.jo, [b]Ibrahimbsoul@yahoo.com

## Abstract

In this article, common experimental techniques and preparation conditions adopted for the synthesis of M-type hexaferrites and their influence on the magnetic properties are briefly reviewed. The effects of various strategies of cationic substitutions on the properties of the hexaferrites are addressed. Further, our synthesis and findings on Co-Ti substituted hexaferrites are presented. It was found that Co-Ti substitution results in improving the saturation magnetization, and reducing the coercivity down to values favorable for high density magnetic recording. Also, evidence of inter-particle interactions in the particulate samples was observed.

## Section Headings

## 1. INTRODUCTION

Permanently magnetizable materials have acquired great scientific, as well as industrial and technological interest due to their crucial role in the fabrication of essential components for a multitude of devices and machines in active use nowadays. The search for new magnetic materials for high performance magnet applications has led to an exponential growth in both the scientific research in this field, and the investment in the development of such materials. This development in materials research was driven by the evolution of new technologies, and the ever increasing demand for the improvement in efficiency, better machine designs, and device miniaturization. The vigorous search for cost-effective magnetic materials had been driven by the large market share of permanent magnet industry, rated at more than $15 billion in 2012, and expected to grow up to over $28 billion by the year 2019 [1].

The discovery of M-type hexagonal ferrites ($MFe_{12}O_{19}$, where M = Ba, Sr, Pb) in early 1950s can be considered one of the most important discoveries in the field of materials development, functionalization, and commercialization during the past few decades. The predominance of these materials in modern magnet industry since 1987 stems from their cost-effectiveness, in addition to their chemical stability, suitability for a wide range of applications, and high performance at relatively high temperatures [2-5]. The basic structural, electrical and magnetic properties of these ferrites are discussed in many previously published works [6-10].

The suitability of a magnetic material for specific applications is mainly determined by the requirements of these applications. For permanent magnet applications, high coercivity and high saturation magnetization are required. For magnetic recording purposes, however, the high coercivity requires an unfavorably high writing filed, which entails high power consumption, and undesirable large device volume. The use of low coercivity material, on the other hand, may involve the risk of losing recorded data as a consequence of the susceptibility of the material to be demagnetized by stray fields. Consequently, a happy balance is called for. This requires modification and tuning the magnetic properties to fit the requirements of low power consumption and device miniaturization on the one hand, and minimize the risk of losing stored information on the other hand. The required value of the coercivity, however, depends on the specific application. While coercivities as low as 300 – 400 Oe could be used for low density magnetic recording applications, higher coercivities in excess of 1000 Oe are required for high density video recording applications [11]. The scenarios for tailoring hexaferrites with modified magnetic properties for specific applications included the adoption of different synthesis techniques to control the grain size and morphology of the produced ceramic, the variations of the stoichiometry of the starting powders and the experimental conditions, and the substitutions for the metal ions in the standard compound. These issues are addressed briefly in the following sections. Further, due to the scarcity of literature concerning the magnetic properties of Co-Ti substituted SrM hexaferrites, we dedicated forthcoming sections to the presentation and discussion of our findings concerning the structural and magnetic properties of these ferrites.

## 2. SYNTHESIS TECHNIQUES FOR THE PREPARATION OF M-TYPE HEXAFERRITES

The conventional ceramic method is widely used for commercial production of hexaferrite powder. This technique, however, could be ineffective in controlling the grain size and morphology of the ferrite powder product. Accordingly, several other techniques involving chemical routes were adopted for the production of the magnet powders. The synthesis of ferrite powders by the various techniques often involved variations of the experimental conditions to determine the optimal conditions for the production of highly pure, high quality powder. Among others, the main factors taken into consideration in synthesizing the ferrite powders included: heat treatment, chemicals used in the starting powders, and stoichiometry of the precursor powder.

In the following subsections, the main techniques used to prepare ferrite precursor powders are briefly discussed. Further details on this matter are found in a previously published review article [2].

### 2.1 CONVENTIONAL CERAMIC METHOD

The standard ceramic method involving mixing and sintering appropriate molar ratios of metal oxides and carbonates precursor powders is a simple method which is widely used in the production of ferrite magnetic materials [12, 13]. Ball milling the starting powders provides smaller particle size and better homogeneity of the resulting ferrite precursor powder, allowing for better crystallization of the ferrite at lower sintering temperatures [2, 14]. Usually, the starting powders (usually $\alpha$-$Fe_2O_3$ and divalent metal (Ba or Sr) carbonate) are mixed and dry-milled or wet-milled by balls in vials made of a mechanically hard material such as stainless steel, zirconia, and tungsten carbide. Wet milling in water, alcohol, or acetone media, for example, improves the milling efficiency. The ball-powder mass ratio is usually in the range 8 – 14, while the powder to liquid ratio is typically 1:1 [8, 15]. The product is then dried, and the resulting dry powder is compressed into desired shapes at high pressure for near full densification (~90% of the theoretical density of 5.28 g/cm$^3$ for BaM and 5.11 g/cm$^3$ for SrM). The compacts are then sintered at high temperature (900° C – 1300° C) for few hours. Compaction of wet-milled BaM powder at a pressure of ~ 250 MPa, and sintering at 1220° C for 3 h resulted in a density of the pellets of about 90% of the theoretical density [16], while compaction at ~ 500 MPa and sintering at 1100° C for 2 h was reported to result in ~ 80 – 93 % densification [17]. On the other hand, pellets compacted at 30 MPa and sintered at 1150° C were reported to have a density below 70% of the theoretical density [18].

The quality of the product was found to depend on the Fe:Ba molar ratio in the starting powder. Molar ratios of Fe:Ba ≥ 12 in the starting materials resulted in the coexistence of unreacted $\alpha$-$Fe_2O_3$ phase with the major M-type phase [19-21]. In a detailed investigation of the effect Fe:Ba ratio on the quality of the synthesized powders, the optimal ratio for the production of a single BaM phase was found to be 11.7 [20]. On the other hand, a detailed and careful study indicated that Ba-rich starting powder mixtures (Fe:Ba < 11) resulted in the coexistence of $BaFe_2O_4$ spinel phase with the major M-type phase at temperatures < 1000° C, and the evolution of more complex oxides such as $Ba_3Fe_2O_6$ at higher temperatures [22].

## 2.2 COPRECIPITATION

This wet chemical technique allows for better reaction of the starting materials at the molecular level, and improves the product quality due to the homogeneity of the metal hydroxides co-precipitated in an aqueous solution of metal salts with the aid of a base (like NaOH) [23]. Appropriate molar ratios of metal chlorides or nitrates are usually dissolved to form homogeneous solutions which are subsequently mixed, and the base is then added drop by drop to promote the precipitation of metal powders. The co-precipitated powders are washed and dried, and the desired M-type hexaferrite is obtained by sintering this dried powder [24]. The required sintering temperature in this method is significantly lower than that required by the conventional solid-state reaction route [25, 26], and the starting solutions normally contain a Fe:Ba ratios lower than the theoretical stoichiometric ratio of 12 [27-30]. In a variation of the method [28], chloride gas was bubbled through a concentrated solution of NaOH to produce NaClO and NaCl solutions to which an appropriate amount of $Fe(NO_3)_3.9H_2O$ was added. Then $BaCl_2.2H_2O$ was added to the deep purple solution at Fe:Ba ratio of 10. The solution was left for 24 h, and then heated at 80° C for 1 h. The solutions were then filtered and rinsed to remove the residual chloride and alcohol. This process leads to the formation of crystalline barium hydroxide, and amorphous ferrihydrite which is structurally related to the hexagonal $\alpha$-$Fe_2O_3$, and thus promotes the evolution of the BaM phase with heat treatment. The results of the study indicated the formation of a pure BaM phase at a low temperature of 800° C.

The effects of the experimental conditions on the quality and properties of the M-type powders prepared by this method were investigated by several workers [30, 31]. The increase of pH from 9 to 13, or the decrease of Fe:Sr ratio in SrM ferrites prepared by coprecipitation was reported to promote the formation of the M-type phase with reduced grain size, and with the minimum coercivity of 4733 Oe and maximum saturation magnetization of 51 emu/g [30]. Further the grain size of BaM hexaferrite with Fe:Ba = 10 was reported to decrease with the increase of pH from 11 to 12.5, while for the samples with Fe:Ba = 10.5 or 11, the grain size increased with increasing pH [31]. The highest saturation magnetization of 66.1 emu/g was reported for the sample with Fe:Ba = 10 prepared at pH = 11 and sintered at 920° C, which decreased to 43.6 emu/g at pH = 12.5. On the other hand, increasing pH resulted in an increase in coercivity from 3400 Oe to 4334 Oe, which is consistent with the decrease of the particle size. The sample with Fe:Ba = 11, however, exhibited the highest coercivity of 4585 Oe at pH = 11, which decreased down to 4435 Oe with increasing pH value up to 12.5. Concurrently, the saturation magnetization of this sample decreased from 60.1 emu/g at pH = 11 to 46.2 emu/g at pH = 12.5.

In a yet another study, it was demonstrated that the saturation magnetization of the coprecipitated powder improved from about 25 emu/g to about 65 emu/g upon increasing the sintering temperature from 640° C to 920° C [32]. Also, the coercivity improved slightly from 5264 Oe at 640° C to 5791 Oe at 920° C, which is indicative of single domain particles with typical high coercivity. Drastically different results, however, were obtained for BaM with Fe:Ba = 12 at pH =12, and sintered at 1300° C, where a saturation magnetization of about 60 emu/g was found, but a rather low remanence of about 15 emu/g and coercivity of 860 Oe were reported [33]. The low coercivity is probably due to the large grain size of several microns in this sample.

## 2.3 SOL-GEL

This method is used to synthesize magnetic powders with controlled particle size distribution. In the standard technique, water solutions of metal chlorides or nitrates are mixed under constant stirring. In the citrate sol-gel method, appropriate molar ratio of citric acid ($C_6H_8O_7$) is then added to the solution under constant stirring. The pH of the solution is adjusted in the range 7 – 9 by addition of a basic solution dropwise with constant stirring [25, 34]. The solution is evaporated at about 80° C, and the resulting highly viscous gel is dried, and sintered to produce the required hexaferrite phase.

Modifications on the technique to improve the quality of the product, and investigate the effects of the experimental conditions were carried out by several investigators. In a variation on the technique, SrM ferrites were prepared by dissolving metal chlorides in citric acid to provide an acidic solution with very low pH, and the powders were subsequently calcined at 1000° C [35]. SrM ferrite prepared by this method exhibited a low saturation magnetization of 30.61 emu/g, and a coercivity of 2213 Oe. On the other hand, in a previous study [36], water solution of $Fe(NO_3)_3$ was precipitated with the aid of ammonia. The precipitate was then dissolved in citric acid with $BaCO_3$, keeping the molar ratio of Fe:Ba = 11.6, and ethylene glycol and benzoic acid were added to the transparent solution. The solution was evaporated at 60° C to produce a highly viscous gel, which was subsequently dried by heating at 170° C. The dried gel was then heat treated in two different ways. In *route a*, gel samples were placed in the furnace, and the heating temperature was set at 1050° C with a heating rate of 4.5° C/min. Different samples were removed from the oven at different temperatures, and quenched to room temperature in air. The samples obtained by this route were annealed at different temperatures for different periods of time. In *route b*, the samples were preheated at 450° C for 5 h prior to heating at temperatures in the range from 500° C to 1050° C for 5 h at the same heating rate. Samples prepared by rout a in the temperature range of 300 – 500° C revealed the formation of γ-$Fe_2O_3$ and $BaCO_3$ phases, and exhibited soft magnetic properties with specific magnetization (measured at 15 kOe) falling in the range 25 – 34 emu/g. The BaM phase developed at temperatures ≥ 550° C, and its fraction in the powder increased with increasing the heating temperature. In addition, α-$Fe_2O_3$ intermediate phase developed in the temperature range of 550 – 900° C, and $BaFe_2O_4$ was observed in the temperature range of 650 - 750° C. The coercivity increased sharply for powders annealed at temperatures above 650° C, reaching about 4200 Oe for powders annealed at temperatures 800 – 900° C. The saturation magnetization for powders heated in this temperature rang also increased up to 65 emu/g, to be compared with 44 emu/g for the sample annealed at 650° C. At temperatures > 900° C, the coercivity decreased, reaching about 3400 Oe at 1050° C, and the specific magnetization increased up to ~ 69 emu/g. The samples prepared by route b exhibited a significant improvement of the magnetic properties in the whole temperature range, and optimum magnetic properties (specific magnetization of 70 emu/g and coercivity of 5950 Oe) were obtained in the range 900 - 950° C [36].

## 2.4 AUTO-COMBUSTION

In this modified citrate sol-gel method, self-propagating combustion of the gel occurs upon heating on a hot plate [37-39]. The starting homogeneous solution is prepared from metal nitrates and citric acid or ethylene glycol fuel with preset molar ratios, and the pH of the solution is adjusted in the range of 7 – 8 by adding a basic solution [40]. The solution is dehydrated by heating at 80° C resulting in a brown/yellow gel, which is subsequently heated at 220 – 240° C to promote auto-combustion yielding foamy powder [41, 42]. The ferrite powder is then obtained by subsequent grinding and sintering the powder at temperatures > 600° C.

Among the experimental variables which influence the properties of the prepared ferrite is the cation-to-fuel ratio and the Fe:Ba ratio [42]. An increase of this ratio from 1:1 to 1:2 was found to improve the magnetic properties, resulting in BaM with saturation magnetization of 55 emu/g, remanence of 28 emu/g, coercivity of 5000 Oe, and energy product of 1.013 MGOe, almost double the magnetic parameters obtained by using a 1:1 ratio [43]. Also, the properties of ferrite powders calcined at 900° C were found to be affected by the Fe:Ba ratio. Samples produced with Fe:Ba = 11 exhibited a saturation magnetization of 51 emu/g and coercivity of 4700 Oe, which improved to about 67 emu/g and 5650 Oe by using Fe:Ba = 9. In addition, the saturation magnetization increased with increasing the calcination temperature from 700° C to 1000° C, and the coercivity increased with increasing temperature up to 900° C, and then decreased down to 4500 Oe at 1000° C due to the presence of large particles growing at high temperatures.

## 2.5 CITRATE PRECURSOR METHOD

This method (also known as the Pecchini method) is used to prepare ultrafine barium ferrite powders at low temperatures via the decomposition of precipitated barium iron citrate complex [2, 44, 45]. The starting solution is prepared by dissolving stoichiometric ratios of Fe and Ba nitrates in deionized water, to which citric acid is added with cation-to-citric acid molar ratio of 1:1. The solution is mixed while adding ammonia drop wise to increase the pH and improve the homogeneity of the solution. The solution is then heated at 80° C to improve the reaction and remove excess ammonia. Then barium iron citrate complex is precipitated by alcohol dehydration through the transfer of the solution drop by drop into ethanol with constant stirring. The yellowish precipitate is subsequently filtered and dried in an oven. The citrate precursor is decomposed at a temperature of 425 – 470° C, resulting ~ 10 nm particles [44, 45]. These particles are normally superparamagnetic at room temperature, and subsequent sintering is required to obtain the BaM powder. Sintering the decomposed powder at 600° C resulted 50 nm particles with saturation magnetization of about 33 emu/g and a coercivity of about 580 Oe, whereas sintering at 700° C resulted in particle growth (to still below 100 nm), and a sharp increase in coercivity up to 4800 Oe, with a slight increase in saturation magnetization to 35 [44]. It is worth mentioning that the saturation magnetization is an intrinsic property of the material, while the coercivity depends on extrinsic parameters such as the particle size. Accordingly, the observed behavior of the magnetic parameters could be an indication that sintering at higher temperatures had the mere effect of increasing the particle size, transforming the powder from superparamagnetic in nature to an assembly of single domain particles with typically high coercivity. Further, it was demonstrated that BaM powder prepared by the citrate precursor method and sintering at 700° C was composed of 60 nm particles with saturation magnetization of 61.5 emu/g, whereas samples fired at 750° C and 800° C exhibited particle growth into the range 80 – 100 nm [46].

## 2.6 HYDROTHERMAL SYNTHESIS

In this method, aqueous solutions of metal nitrates are coprecipitated with a strong base solution such as NaOH or KOH with the appropriate $OH^-$:$NO_3^-$ and Fe:Ba ratios. The solution containing the metallic precipitates are then hydrothermally treated in an autoclave at temperatures in the range 150 – 290° C. The resulting particles are then filtered, washed and dried in an oven. To improve the magnetic characteristics of the product, the dried powder is sintered at temperatures of 1100 - 1200° C.

The experimental conditions including the hydroxide-nitrate ratio, the Fe:Ba ratio, the hydrothermal heat treatment temperature and duration, and the sintering temperature of the powder were found to be critical parameters for the properties of the final product [47]. Normally, hydroxide to nitrate ratio > 2 was found to be necessary for the formation of BaM, where lower ratios resulted in presence of intermediate iron oxide phase [47, 48]. However, hydrothermal treatment of aqueous solutions with much higher hydroxide-nitrate ratios (16) at temperature 150° C resulted in the formation of superparamagnetic particles ~ 10 nm in diameter [49, 50]. Further, the duration of the hydrothermal treatment was found critical for the formation of M-type phase. For example, samples prepared with Fe:Ba = 8 and $OH^-:NO_3^-$ = 2 at 230° C were found to contain $\alpha$-$Fe_2O_3$ as a major phase for heat treatment periods ≤ 10 h, whereas BaM phase was formed in the sample heat treated for 25 h [47].

The Fe:Ba ratio was also found to be crucial to the formation of the M-type using hydrothermal synthesis. A ratio < 10 was found necessary for the formation of BaM phase in samples prepared at a reaction temperature of 230° C for 48 h with fixed hydroxide-nitrate ratio of 2 phase, where higher values resulted in the predominance of iron oxide in the product [47].

## 2.7 MOLTEN SALT METHOD

This procedure is used to synthesize M-type hexaferrites with large crystals. The basic procedure involves mixing the reactants (barium carbonate and iron oxide precursor powders) with a solvent consisting of NaCl-KCl salt mixture, and heating the reaction mixture at 800 - 1100° C [51]. The procedure results in a dry cake in which the magnetic hexaferrite powder is entrapped. The dried body is then crushed and washed with distilled water to remove the salt, and the magnetic powder is obtained. Large variations of the magnetic properties were revealed depending on experimental conditions including the starting reactants, solvent composition and purity, and the heat treatment [51, 52]. Hexagonal plates of BaM with basal dimension < 1.5 μm and optimal magnetic properties (saturation magnetization of 72 emu/g, and $H_c$ = 4300 Oe) were synthesized under optimal experimental conditions [51].

In a variation to the method, BaM powder is first prepared by coprecipitation, and the coprecipitated particles are mixed with KCl flux at a BaM to salt weight ratio of 1:1 [53]. BaM particles were synthesized by initially heating the mixture at 450° C, and then at 950° C. The product was then washed with deionized water to remove the salts, and dried at 80° C in an oven.

## 2.8 GLASS CRYSTALLIZATION

In this method, the ferrite powder is mixed and melted with a glass matrix. Subsequently, the melt is rapidly quenched to produce an amorphous matrix containing the ferrite. The crystallization of the hexaferrite phase is achieved by annealing at temperatures > 600° C, and the magnetic ferrite is retrieved by dissolving the amorphous matrix in a dilute acid, which does not dissolve the hexaferrite phase [54, 55]. Samples prepared by this method at different heat treatments revealed coercivity ranging from 2600 Oe to 5350 Oe [56].

## 2.9 SPRAY PYROLYSIS

A system consisting of an ultrasonic droplet generator, a quartz high temperature reactor, and a powder collector was used for the preparation of hexaferrite [57-59]. The droplets of the precursor solution are injected into the reactor by a gas flow, which can be adjusted for optimal results. The evaporation of the droplets 900° C resulted in a powder consisting of spherical particles. Improvement of the magnetic properties was achieved by subsequent heat treatment. As expected, the post heat treatment influenced the particle size and coercivity greatly, and the sample heat-treated at 800° C revealed a high coercivity of 6000 Oe [60].

## 2.10 FURTHER DEVELOPMENTS IN SYNTHESIS ROUTES

The pursuit of high quality hexaferrite materials with high magnetic properties have led to the development of a variety of powder synthesis routes, some of which were discussed in the previous subsections. The final product, depending on the stoichiometry of the starting powder or solution mixtures and the adopted experimental conditions, often resulted in secondary phases which influenced the magnetic properties negatively. Accordingly, various remedies were proposed to improve the purity of the ferrite powder, and enhance its magnetic properties. These include modifications of the prevailing experimental conditions, and combining synthesis routes, as well as proposing new synthesis methods.

Barium hexaferrite powder prepared by the conventional ceramic method and calcined at 900° C was reported to be composed of a mixture of equilibrium phases, consisting of BaM and the intermediate $BaFe_2O_4$ and $Fe_2O_3$ phases [61]. Small additions of $B_2O_3$, however, was found to result in a single BaM phase with enhanced remanent magnetization at such low sintering temperature. Specifically, the sample with 1 wt.% $B_2O_3$ addition revealed remanence magnetization of 28 emu/g, saturation magnetization of 54 emu/g, and coercivity between 2000 and 3000 Oe; these properties are suitable for magnetic recording applications. Further improvements of the magnetic properties were achieved with $B_2O_3$ addition and etching with diluted HCl solution [62]. Specifically, the sample with 0.1 wt. % $B_2O_3$ exhibited a rather high remanent magnetization of 34.9 emu/g and magnetization of 63.3 emu/g measured at 1.5 T applied field. The coercivity, however, did not exhibit a systematic behavior with HCL washing, and the enhancement of the remanence and saturation magnetization were attributed to removal of nonmagnetic impurity phases which may not be detectable by XRD.

The coercivity of BaM magnets prepared by ball milling and sintering was also found to be enhanced by small additions of $V_2O_3$ due to the formation of finer powders with smaller average particle size and narrower particle size distribution [63]. At low $V_2O_3$ concentrations (0.7 wt.%) BaM samples sintered at 1100° C reveled an increase of the coercivity from 3.5 kOe to 4.1 kOe, while the saturation magnetization and remanent magnetization remained almost the same as for the un-doped sample (68.6 emu/g and 37.4 emu/g, respectively). At 3.5 wt.% $V_2O_3$, however, significant fractions of nonmagnetic secondary phases were observed, and the saturation magnetization and remanent magnetization reduced dramatically down to 21.8 emu/g and 11.4 emu/g, respectively. Higher sintering temperatures improved the saturation magnetization (26.6 emu/g at 1300° C), and reduced the coercivity down to 1.6 kOe. The saturation magnetization and remanence were improved significantly (to 59.6 emu/g and 32.8 emu/g, respectively) by adding an extra 14 wt.% $BaCO_3$ (for (Fe+V):Ba = 6.3), sintering at 1200° C, and washing with diluted HCl solution. The coercivity of the product was 2.1 kOe, and the relatively high remanent magnetization makes the product suitable for magnetic recording media.

The oxalate precursor route adopted for the synthesis of BaM ferrite was found to produce a powder with a rather low coercivity [64]. In this method, metal chlorides with Fe:Ba = 12 were dissolved in equal amounts of oxalic acid, and mixed with continuous stirring using a magnetic stirrer for 15 min. The solution was then heated at 80° C with constant stirring, and dried at 100° C overnight. The dried powder precursors were then annealed at temperatures in the range of 800 – 1200° C for two h. The prepared powders exhibited an increase of both the saturation magnetization and coercivity with increasing the annealing temperature from 900 to 1100° C. The maximum saturation magnetization was found to be 66.36 emu/g, while the maximum coercivity was only about 640 Oe.

The direct crystallization of BaM ferrite from an aerosol was realized for the first time by a combination of pyrolysis and citrate precursor methods [65]. The barium iron citrate precursor solution with Fe:Ba = 12 was nebulized at a flow rate of 0.5 ml/min to produce a stream of fine droplets, which was passed through the low temperature furnace (200 – 250° C) of the reactor to evaporate the solvent. The dried barium iron citrate particles were subsequently passed through a high temperature furnace, resulting in a powder consisting of submicron hollow spheres with a rather low saturation magnetization of 5.6 emu/g and a coercivity of 2500 Oe. Further heat treatment of the powder at 1000° C improved the magnetic properties significantly, where the saturation magnetization increased up to 50.0 emu/g and the coercivity to 5600 Oe. When the powders were hand milled to eliminate the effects of particle aggregation, the coercivity of the heat-treated sample further increased to 5900 Oe, but the saturation decreased to 42.6 emu/g. The obtained coercivity by this synthesis route is probably one of the highest reported for pure BaM ferrite.

In addition, a new synthesis route, named ammonium nitrate melt technique (ANMT) [66], was proposed for the production of high quality BaM ferrite using low Fe:Ba ratios [67]. In this method, the desired proportions of $BaCO_3$ and $Fe_2O_3$ powders were mixed with the ammonium nitrate melt and stirred with a magnetic stirrer. The resulting thick solution was evaporated at 260° C, resulting a reddish precipitate which was subsequently preheated at 450° C for 5 h. Parts of the resulting powder were then heat treated at temperatures in the range of 800 – 1200° C to identify the optimal heat treatment. The results of the study revealed that the powder with Fe:Ba ratio of 2 gives the highest saturation magnetization. When this powder was sintered at 1100° C, a mixture of $BaFe_2O_4$ and BaM phases was observed. Upon washing the powder with HCl solution, the $BaFe_2O_4$ phase disappeared, and a single BaM phase with saturation magnetization of 66.7 emu/g, remanent magnetization of 38.5 emu/g, and coercivity of 4228 Oe was obtained.

The above discussion indicates that the synthesis method and the experimental conditions play a critical role in the quality and magnetic properties of the produced hexaferrite powder. Hexaferrites with high coercivity and saturation magnetization are suitable for a wide variety of permanent magnet (PM) applications, including low-power motors in home appliances, motors and switches in auto-industry, loud speakers, current meters, magnetic seals, magnetic shielding, toys, etc. Ferrites with lower coercivities, however, are suitable for magnetic recording applications. While materials with coercivities in the range of few hundred Oe can be used for low density longitudinal magnetic recording (LLMR) media, materials with higher coercivities up to 1200 Oe can be used for high density longitudinal recording media (HLMR). Materials with higher coercivity are not suitable for longitudinal recording, and can effectively be used for high density perpendicular recording (HPMR) applications. The production of a hexaferrite material with the desired properties requires optimization of the experimental procedures to produce high quality single-domain powders with controlled particle size. In Table 1, some of the experimental findings concerning the magnetic properties of M-type hexaferrites with potential permanent magnet and magnetic recording applications are presented.

*Table 1: Magnetic properties of M-type hexaferrites prepared by different techniques.*

| Synthesis Method | $M_s$ (emu/g) | $M_r$ (emu/g) | $H_c$ (Oe) | Application | Reference |
|---|---|---|---|---|---|
| Conventional Solid State Method | 61 | 32 | 2080 | HPMR | [68] |
| | 49 | 24 | 1005 | HLMR | [69] |
| | 57 | | 1943 | HPMR | [70] |
| | 72 | | 4200 | PM | [71] |
| | 71 | 37 | 4020 | PM | [72] |
| | 60 | | 4000 | PM | [73] |
| Coprecipitation | 64 | 31 | 4700 | PM | [74] |
| | 72 | | 5340 | PM | [75] |
| | 69 | 36 | 5440 | PM | [76] |
| | 65 | | 5540 | PM | [77] |
| | 70 | | 5044 | PM | [25] |
| | 60 | 15 | 860 | HLMR | [33] |
| | 71 | | 6400 | PM | [78] |
| | 65 | 36 | 5791 | PM | [32] |
| Sol-gel | 61 | | 1827 | HPMR | [79] |
| | 49 | | 4800 | PM | [34] |
| | 61 | 37 | 4996 | PM | [80] |
| | 59 | 36 | 1920 | HPMR | [81] |
| | 61 | 36 | 5692 | PM | [82] |
| | 70 | | 5900 | PM | [83] |
| | 70 | | 5950 | PM | [36] |
| | 67 | | 5650 | PM | [42] |
| Auto combustion | 55 | 28 | 5000 | PM | [43] |
| | 51 | 36 | 2037 | HPMR | [40] |
| | 64 | | 750 | HLMR | [84] |
| | 50 | 31 | 5017 | PM | [85] |
| | 60 | | 4250 | PM | [86] |
| | 40 | 22 | 5689 | PM | [87] |
| | 74 | | 5163 | PM | [88] |
| Hydrothermal | 40 | | 2500 | PM, HPMR | [49] |
| | 59 | 20 | 1350 | HPMR | [89] |
| | 64 | | 2300 | PM, HPMR | [47] |
| Molten salt | 59 | | 4820 | PM | [90] |
| | 72 | | 4650 | PM | [53] |
| | 72 | | 4300 | PM | [51] |

## 3. EFFECTS OF CATIONIC SUBSTITUTION

In the preceding section, we addressed the effects of synthesis method, and experimental conditions on the magnetic properties of hexaferrites. It was demonstrated that great variations of the magnetic properties were achieved through modifications of the experimental procedures. However, such variations were mainly due to the quality and purity of the produced hexaferrite, and the particle size and morphology of the powder. An efficient scenario for the modification of the magnetic properties can also be achieved by cationic substitution in the standard M-type hexaferrite, which normally result in modifying the intrinsic properties such as the magnetocrystalline anisotropy [91, 92].

Cationic substitution in M-type hexaferrite could involve the divalent ($M^{2+}$) ions, and/or the $Fe^{3+}$ ions. Normally, Sr is used to replace Ba in BaM-type hexaferrite [59, 93, 94], where SrM hexaferrite with improved coercivity of ≥ 6.4 kOe was prepared by different experimental procedures [95, 96]. The use of $Pb^{2+}$ and $Ca^{2+}$, on the other hand, generally result in lowering the coercivity of M-type hexaferrite [97-99]. In an earlier publication, however, the addition of CaO was reported to improve the coercivity of SrM ferrite [100]. Further, the effect of Ca and Ca-Sr substitution for Ba in BaM ferrite prepared by the conventional ceramic method, pre-sintering at 600° C, and sintering at 1100° C was investigated [101]. The results of the study revealed an increase in coercivity from 2.75 kOe for BaM to 3.2 kOe for $Ba_{0.5}Ca_{0.5}Fe_{12}O_{19}$, with a concurrent decrease in saturation magnetization from 53.04 emu/g to 33.17 emu/g.  The sample $Ba_{0.5}Ca_{0.25}Sr_{0.25}Fe_{12}O_{19}$, on the other hand, exhibited a higher saturation magnetization of 50.20 emu/g, and a lower coercivity of 2.98 kOe.

The effects of the substitution of the divalent metal ions by rare-earth ions was also given some attention. The substitution of Ba by Eu in BaM prepared sol–gel method was found to result in an increase the coercivity from 1.92 kOe for the un-doped sample up to 6.12 kOe for $Ba_{0.75}Eu_{0.25}Fe_{12}O_{19}$ [81]. The saturation magnetization at this stoichiometry dropped down from 59.21 emu/g to 45.14 emu/g, while the remanent magnetization decreased slightly from 35.66 emu/g to 34.52 emu/g. Also, it was found that only 10% substitution of Ba by La in BaM prepared by a reverse micro-emulsion route leads to high magnetic properties of $M_s$ = 66.5 emu/g, $M_r$ = 34.6 emu/g, and $H_c$ = 5229 Oe [102]. In addition, BaM nanofibers prepared by citrate sol–gel method and electrospinning, followed by heat treatment, revealed an increase in saturation magnetization from about 64.5 emu/g up to 77.2 emu/g, and a decrease in coercivity from 4323 Oe down to 3565 Oe at 5% La substitution for Ba [103]. At 10% La substitution, the saturation magnetization dropped down to 71.0 emu/g, and the coercivity increased to about 4160 Oe. The results of this study also revealed an enhancement of the microwave absorption properties with La substitution for Ba, and the La-doped compounds were reported to have a significant potential for microwave absorption applications. On the other hand, the substitution of Sr by La–Ce combination in Zn-substituted SrM ferrite prepared by the conventional solid state route was reported to induce a reduction in both the saturation magnetization and coercivity of the hexaferrites calcined in air at 1200° C [104]. The saturation magnetization of the calcined powders was enhanced by annealing at 1100° C in $N_2$ environment, and a product with saturation magnetization as high as 77.4 emu/g and coercivity of 2343 Oe was obtained for $Sr_{0.7}La_{0.1}Ce_{0.2}Fe_{11.7}Zn_{0.3}O_{19}$ ferrite. Also, in a recent publication, it was demonstrated that the saturation magnetization improved upon increasing *x* up to 0.2, and the coercivity increased with increasing *x* up to 0.3 in $Sr_{1-x}La_xFe_{12-x}Co_xO_{19}$ [105]. Further, partial substitution of Ba by either La or Pr in BaM powders prepared by auto-combustion was reported to result in an improvement of the saturation magnetization and coercivity of the ferrite, Pr substitution being more effective, especially in increasing the coercivity [106].

The substitution of $Fe^{3+}$ by a trivalent metal, or by combinations of divalent and tetravalent metals was extensively investigated with the purposes of modifying the magnetic properties to fit specific applications. The effects of such substitutions will be the subject of the following subsections.

## 3.1 $Fe^{3+}$ SUBSTITUTION BY TRIVALENT METAL IONS

In general, the substitution of $Fe^{3+}$ by a trivalent metal such as Al, Cr, or Ga results in an increase in coercivity, with Al being the most effective [68, 71, 107-109]. Since the coercivity is influenced by the particle size, a reduction in coercivity is normally observed at elevated sintering temperatures due to particle growth beyond the critical single domain size [110]. Since the early times of BaM synthesis and characterization, small additions of kaolin ($Al_2O_3(SiO_2).2H_2O$), however, were reported to inhibit particle growth with increasing the sintering temperature, and improve the magnetic properties [13]. The improvement in the maximum energy product $(BH)_{max}$ with kaolin addition, however, was not significant, where it ranged from 3.34 MGOe with 1 wt.% kaolin to 2.55 MGOe with 3 wt.% kaolin addition; this is to be compared with values in the range of 2.97-3.52 MGOe for BaM with no kaolin addition and sintering at different temperatures [13]. This is possibly due to the competing effects of increasing the coercivity and decreasing the saturation magnetization with kaolin addition.

$BaFe_{10}Al_2O_{19}$ prepared by ball milling and sintering at 1100° C revealed an increase in coercivity up to 9.3 kOe, accompanied by a significant reduction of the saturation magnetization down to 36 emu/g [108]. Also, a coercivity of 7.1 kOe and a saturation magnetization of 21.6 emu/g were reported for $SrFe_{10}Al_2O_{19}$ ferrite prepared by auto-combustion and calcination at 950° C, and a record coercivity of 16.2 kOe with a rather low saturation magnetization of 11.80 emu/g were reported for $SrFe_8Al_4O_{19}$ [111]. In addition, an increase in coercivity up to 7.4 kOe, and a reduction in saturation magnetization down to 36.50 emu/g were observed in $SrFe_{10}Al_2O_{19}$ prepared by auto-combustion and calcination at 1100° C [112]. Further, $SrFe_{10.7}Al_{1.3}O_{19}$ prepared by glass crystallization and calcination at 950° C revealed a high coercivity of 10.18 kOe, accompanied by a reduction in saturation magnetization down to 18.4 emu/g [113]. Furthermore, a relatively high saturation magnetization of 50.5 emu/g with coercivity of 6.0 kOe was obtained in $BaFe_{11.2}Al_{0.8}O_{19}$ prepared by ball milling and calcination at 1100° C [71]. However, contradictory results on Al-doped BaM were reported elsewhere [114].

The coercivity of Al-substituted SrM ferrites was improved by partial substitution of Sr by a rare-earth (RE) element. A systematic study of the effect of the type of RE ion substitution on the magnetic properties of $Sr_{0.9}RE_{0.1}Fe_{10}Al_2O_{19}$ revealed a significant improvement of the coercivity of SrM ferrite, Pr being the most effective in enhancing the coercivity from 7.4 kO up to 11.0 kOe [112]. The saturation magnetization, on the other hand, decreased from 36.5 emu/g down to 30.8 emu/g with Pr substitution. At this point, it is worth mentioning that although $Al^{3+}$ substitution resulted in a significant increase in coercivity, the effect of such substitution leads to deterioration of the saturation magnetization, and limits the applicability of the product to applications demanding high coercivity, without necessarily very high saturation magnetization values. Such applications include the production of permanent magnets for devices operating for long times in environments of high stray fields.

$Cr^{3+}$ substitution for $Fe^{3+}$ was also reported to enhance the coercivity of M-type hexaferrite [115, 116]. Cr-substitution for Fe in BaM ferrite ($BaFe_{12-x}Cr_xO_{19}$) prepared by sol–gel auto-combustion method and calcination at 1100° C revealed a decrease in saturation magnetization, and an increase in coercivity with increasing Cr content up to $x = 0.8$ [116]. The decrease in saturation magnetization was attributed to magnetic dilution or spin canting as a result of Cr substitution for Fe at 2a and 12k sites of the hexaferrite lattice. The increase in coercivity from 2.0 kOe at $x = 0$ up to 5.2 kOe at $x = 0.8$, on the other hand, was associated with the reduction in grain size with increasing Cr content. Similar effects on grain size and morphology, and on the magnetic properties of Cr substitution for Fe in SrM ferrite ($SrFe_{12-x}Cr_xO_{19}$) prepared by microwave hydrothermal route and calcination at 950° C were observed [117]. In this latter study, the remanent magnetization decreased from 34 emu/g to 24 emu/g, the average grain size reduced from 660 nm to 280 nm, and the coercivity increased from 3291 Oe to 7335 Oe upon increasing $x$ from 0.1 to 0.9. The increase in coercivity with increasing Cr content was associated with the reduction of the grain size, and the increase of the nonmagnetic $\alpha$-$Fe_2O_3$ phase fraction, which acts as pinning centers for domain wall motion.

The substitution of $Fe^{3+}$ by $Gd^{3+}$ [118] or $Ce^{3+}$ [119] was also found to improve the saturation magnetization and the coercivity of BaM ferrite. Also, the addition of 3 wt. % $Bi_2O_3$ to SrM hexaferrite sintered at 900° C was found to improve both the saturation magnetization and coercivity by more than double the values for the sample with no additives [120].

## 3.2 $Fe^{3+}$ SUBSTITUTION BY DIVALENT-TETRAVALENT METAL ION COMBINATIONS

For a wide range of applications, high saturation magnetization is required, but not necessarily very high coercivity. The use of M-type hexaferrites for applications such as high density magnetic recording media requires reduction of the coercivity to $\gtrsim$ 1000 Oe without reducing the remanent magnetization, which could be satisfactory at the lower end of about 20 emu/g [121, 122]. Such properties could be achieved by special substitutions for Fe in M-type hexaferrite. Accordingly, in an effort to understand the nature of magnetic interactions, and modify the magnetic properties of hexaferrites, extensive research work had been carried out on the synthesis and characterization of M-type hexaferrites with $Fe^{3+}$ ions substituted by special combinations of divalent–tetravalent metal ions [73, 123-142].

Specifically, Co–Zn–Nb substitution in BaM ferrite prepared by glass crystallization was found to reduce the coercivity to the range of 500 – 2000 Oe, leaving the saturation magnetization relatively high for high density magnetic recording applications [143]. Also, it was found that Zn–Nb substitution offers an advantage over Co–Nb and Co–Ti substitutions by reducing the switching field distribution for a better high density recording performance. In addition, Co–Sn substitution in BaM ($BaCo_xSn_xFe_{12-2x}$) prepared by a reverse micro-emulsion technique resulted in a gradual decrease of the saturation magnetization, the remanent magnetization, and coercivity with increasing $x$ [102]. However, the saturation magnetization obtained by this method was found superior to that obtained by other chemical methods. The results of the study indicated that at $x$ = 0.5, $M_s$ = 70.4 emu/g, $M_r$ = 34.0 emu/g, and $H_c$ = 1510 Oe; these properties are suitable for high density magnetic recording.

Further, the effects on the magnetic properties of A–Sn (A = Co, Ni, Zn) substituted BaM ($BaFe_{12-2x}Sn_xA_xO_{19}$) were investigated as a function of heat treatment, substitution level, and preparation method [14]. The samples prepared by solid state reaction (SSR) were fired at 1300° C, whereas those prepared by high energy ball milling (HEM) and chemical coprecipitation (CC) methods were fired at 1000 – 1100° C and 750° C, respectively. The saturation magnetization of the Co – Sn and Ni – Sn substituted samples prepared by SSR route dropped sharply from 60 – 65 emu/g down to 20 – 25 emu/g as $x$ increased from 0.5 to 1.0. The Zn – Sn substituted samples prepared by the same method, however, exhibited a small increase from ~ 53 emu/g at $x$ = 0.1 to ~ 57 emu/g at $x$ = 1.0, and then dropped sharply down to ~ 20 emu/g at $x$ = 1.5. The samples with $x$ = 1.0 prepared by HEM and CC methods exhibited higher saturation magnetizations than those prepared by SSR method, and these variations were attributed to the modification of the preferential site occupation of the substituents at this substitution level. The coercivity of the Co – Sn and Ni – Sn substituted samples prepared by SSR decreased sharply from ~ 1100 – 1200 Oe at $x$ = 0.1 to ~ 500 – 700 Oe at $x$ = 0.5, and then decreased at a slower rate at higher concentrations of the substituents. The Zn – Sn substituted samples, however, exhibited significantly lower coercivities in the whole concentration range. The results of the study indicated that the CC method resulted in higher coercivities, where at $x = 0.1$ the coercivities of all compounds were in the range 4140 – 4380 Oe, characteristic of hard ferrite magnets appropriate for permanent magnet applications due to their relatively high saturation magnetization. On the other hand, at $x = 1.0$ the coercivities of the Co – Sn and Zn – Sn samples were 1040 Oe and 1180 Oe, respectively, and their saturation magnetization in the range 40 – 45 emu/g makes them suitable for high density magnetic recording applications. At this substitution level, however, the Ni – Sn substituted sample exhibited weaker magnetic properties with a significantly lower coercivity of 490 Oe, and a saturation magnetization below 30 emu/g.

In particular, Co–Ti substitution received a considerable interest due to the early realization of the effectiveness of this combination in reducing the coercivity down to values suitable for applications such as magnetic recording and data storage media [11, 54, 139, 144-155]. Materials with relatively high saturation magnetization and coercivities up to 1200 Oe are suitable for normal longitudinal magnetic recording, where the higher end of the coercivity is used for high density recording. Materials with higher coercivities, however, are not suitable for longitudinal recording, and could be used for high density perpendicular magnetic recording media [151, 154]. Further, potential applications of Co – Ti substituted M-type ferrites in multi-layer chip beads and other microwave applications in the hyper-frequency range were reported [156-159]. However, the work devoted to the synthesis and investigation of the magnetic properties of Co–Ti substituted SrM ferrites was extremely limited in the literature. We therefore devote the remaining of this article to our results concerning the synthesis and characterization of $SrFe_{12-2x}Co_xTi_xO_{19}$ ferrites prepared by ball milling and calcination.

# 4. Co-Ti SUBSTITUTED SrM FERRITE

## 4.1 EXPERIMENTAL PROCEDURES

$SrFe_{12-2x}Co_xTi_xO_{19}$ ($x$ = 0, 0.2, 0.4, 0.6, 0.8 and 1.0) precursor powders were prepared from spec pure $SrCO_3$, $Fe_2O_3$, $TiO_2$ and CoO powders (Sigma Aldrich-make). Each sample was prepared by ball-milling 8g of the starting powder mixture using a planetary ball-mill (Fritsch Pulverisette-7) with balls (10 cm in diameter) and cylindrical vial (50 $cm^3$) of hardened steel. The milling was carried out at 250 rpm for 16 h with ball to powder mass ratio of 8:1. The as-milled powders were annealed in air atmosphere at 1100°C for 2 h. X-ray diffraction (XRD) analysis was carried out in Philips X'Pert PRO X-ray diffractometer (PW3040/60) with $CuK_\alpha$ radiation ($\lambda$ = 1.5405 Å) . XRD patterns for all fabricated samples were refined based on Rietveld analysis using FullProf suite 2000 software [160]. The particle size and morphology of the prepared samples were examined by scanning electron microscopy (SEM) (FEI Quanta 600). The magnetic measurements were carried out using vibrating sample magnetometer (VSM) (MicroMag 3900, Princeton Measurements Corporation), with a maximum applied field of 10 kOe.

## 4.2 XRD ANALYSIS

Fig. 1 shows XRD patterns of Co–Ti doped strontium ferrites ($SrFe_{12-2x}Co_xTi_xO_{19}$) with different doping concentrations, together with the standard pattern for $SrFe_{12}O_{19}$ (JCPDS file no: 033-1340). XRD patterns of the samples with $x$ up to 0.8 revealed pure SrM phase ($SrFe_{12}O_{19}$) with space group $P6_3/mmc$, while a small peak at $2\theta = 33.1°$ belonging to a minority $\alpha$-$Fe_2O_3$ phase appeared in the sample with $x = 1.0$. Therefore we may conclude that Co and Ti ions with concentrations in the range reported in this work substituted $Fe^{3+}$ ions in the $SrFe_{12}O_{19}$ lattice without affecting its hexagonal structure appreciably.

Rietveld structural refinements were carried out on XRD patterns of all fabricated samples, and the refined lattice constants and goodness of fit (indicated by $\chi^2$ value) are shown in Table 2. A representative refined pattern is shown in Fig. 2 for the pure sample ($x = 0.0$). The relatively low values of $\chi^2$ indicate good fits with reliable structural parameters. The lattice parameters for the sample with $x = 0.0$ agree well with those reported for strontium hexaferrite [161-163]. Also, the lattice parameters of the Co–Ti substituted samples did not change appreciably in the whole range of substitution level. These results clearly indicate that doping of strontium ferrite with Co and Ti did not induce noticeable structural changes in SrM hexaferrite.

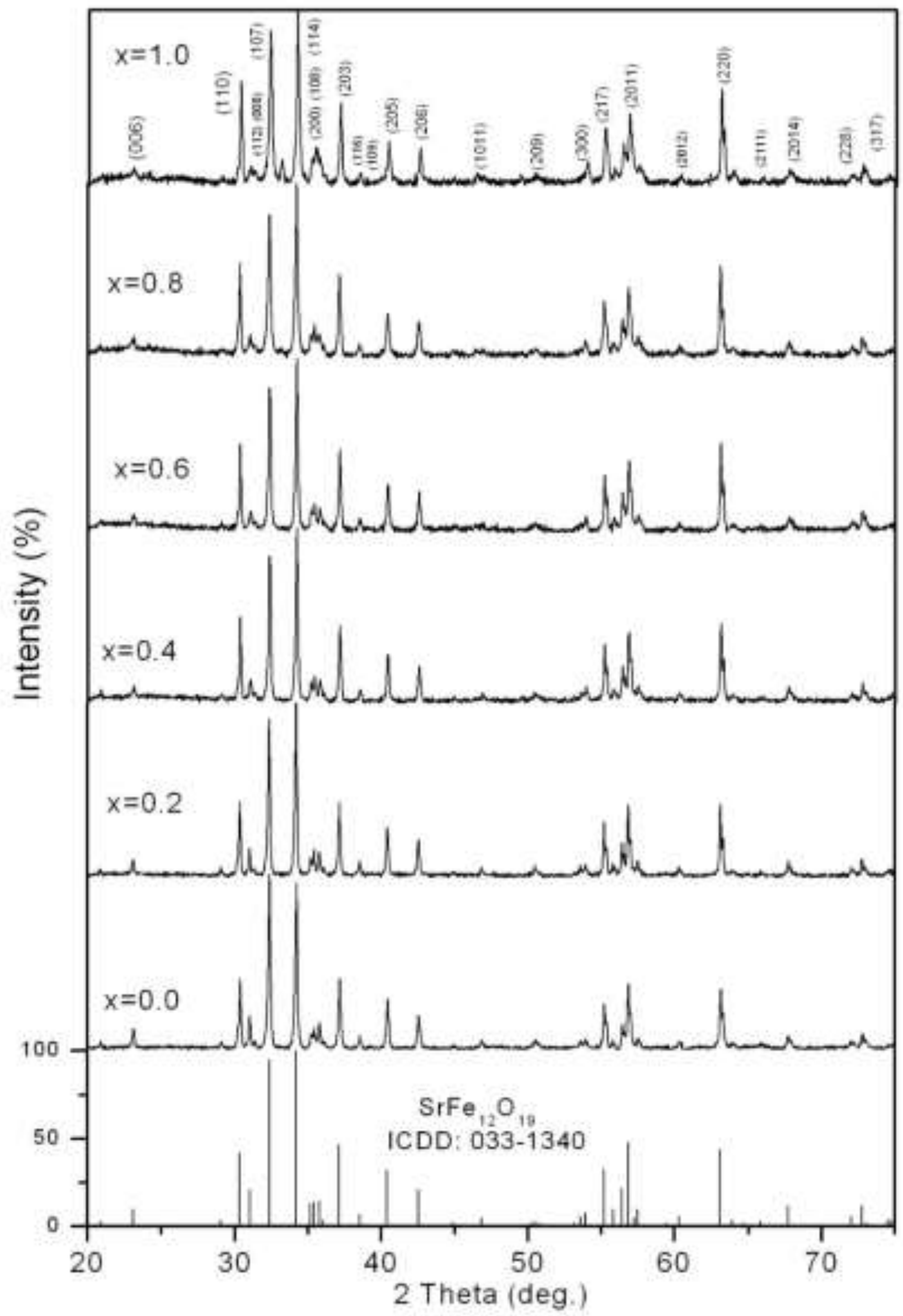

Fig. 1. XRD patterns of $SrFe_{12-2x}Co_xTi_xO_{19}$ ($x$ = 0.0 to 1.0).

*Table 2. Refined structural parameters and crystallite size of Co-Ti substituted SrM ferrites.*

| $x$ | $\chi^2$ | $a = b$ (Å) | $c$ (Å) | $V$ (Å)$^3$ | Average crystallite size ($D_{XRD}$) nm |
|---|---|---|---|---|---|
| 0.0 | 1.08 | 5.8814 | 23.0355 | 690.0652 | 57 |
| 0.2 | 0.90 | 5.8802 | 23.0229 | 689.4157 | 60 |
| 0.4 | 0.94 | 5.8809 | 23.0192 | 689.4473 | 57 |
| 0.6 | 1.01 | 5.8813 | 23.0170 | 689.4847 | 55 |
| 0.8 | 1.33 | 5.8826 | 23.0226 | 689.9492 | 49 |
| 1.0 | 1.63 | 5.8828 | 23.0254 | 690.0911 | 45 |

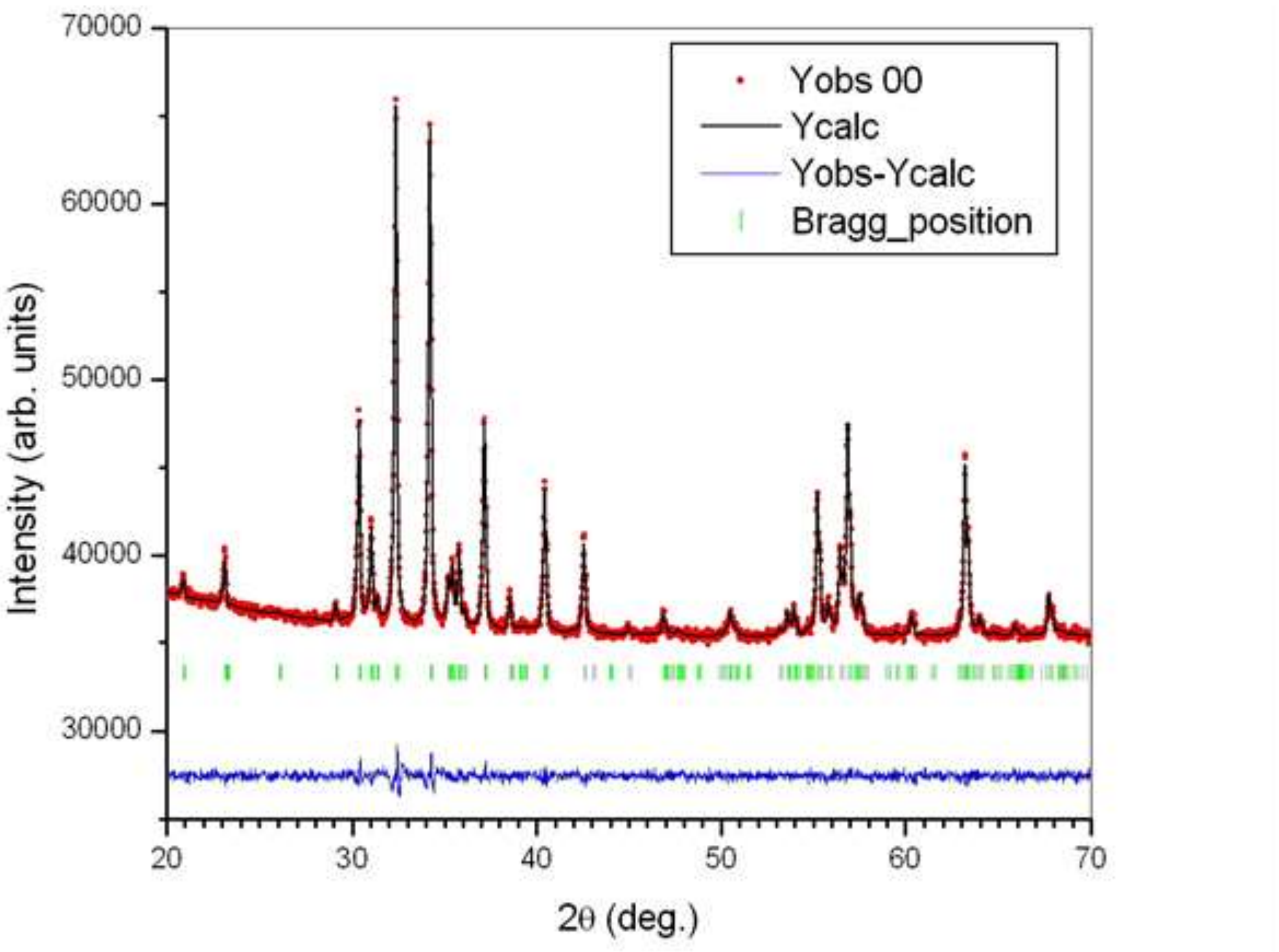

Fig. 2. Refined XRD pattern for $SrFe_{12}O_{19}$.

The average crystallite size was determined from the (114) and (107) reflections using the method of Stokes and Wilson, where the coherence length along the direction perpendicular to the reflecting plane is given by [164]:

$$D = \lambda/(\beta \cos \theta) \qquad (1)$$

where $D$ is the crystallite size, $\beta$ the integral peak breadth defined as the ratio of the integrated intensity to the maximum intensity of the peak, $\lambda$ the wavelength of radiation (1.54056 Å), and $\theta$ the peak position. The average crystallite size (Table 2) in the samples with $x$ up to 0.6 remain in the narrow range of 55 – 60 nm, and only higher substitution levels show tendency to reduce the crystallite size, which dropped down to 45 nm for the sample with $x$ = 1.0.

### 4.3 SEM IMAGING

The grain size and morphology for all samples were examined by SEM imaging. Fig. 3 shows representative SEM photographs of Co–Ti doped strontium ferrite powders with doping concentration of 0.0, 0.2, 0.4 and 0.8. Generally speaking, all samples are composed of agglomerated platelet-like grains with morphology that did not change noticeably with Co–Ti substitution. The grain size, however, revealed small decrease with substitution, and all samples are composed of grains < 1 μm in dimension, which is within the critical single domain size [2]. There the image of the pure sample ($x$ = 0.0) revealed grain size generally in the range from 200 nm to 0.8 μm, while the images of the Co–Ti substituted indicated grain sized ranging from 150 nm to 0.5 μm.

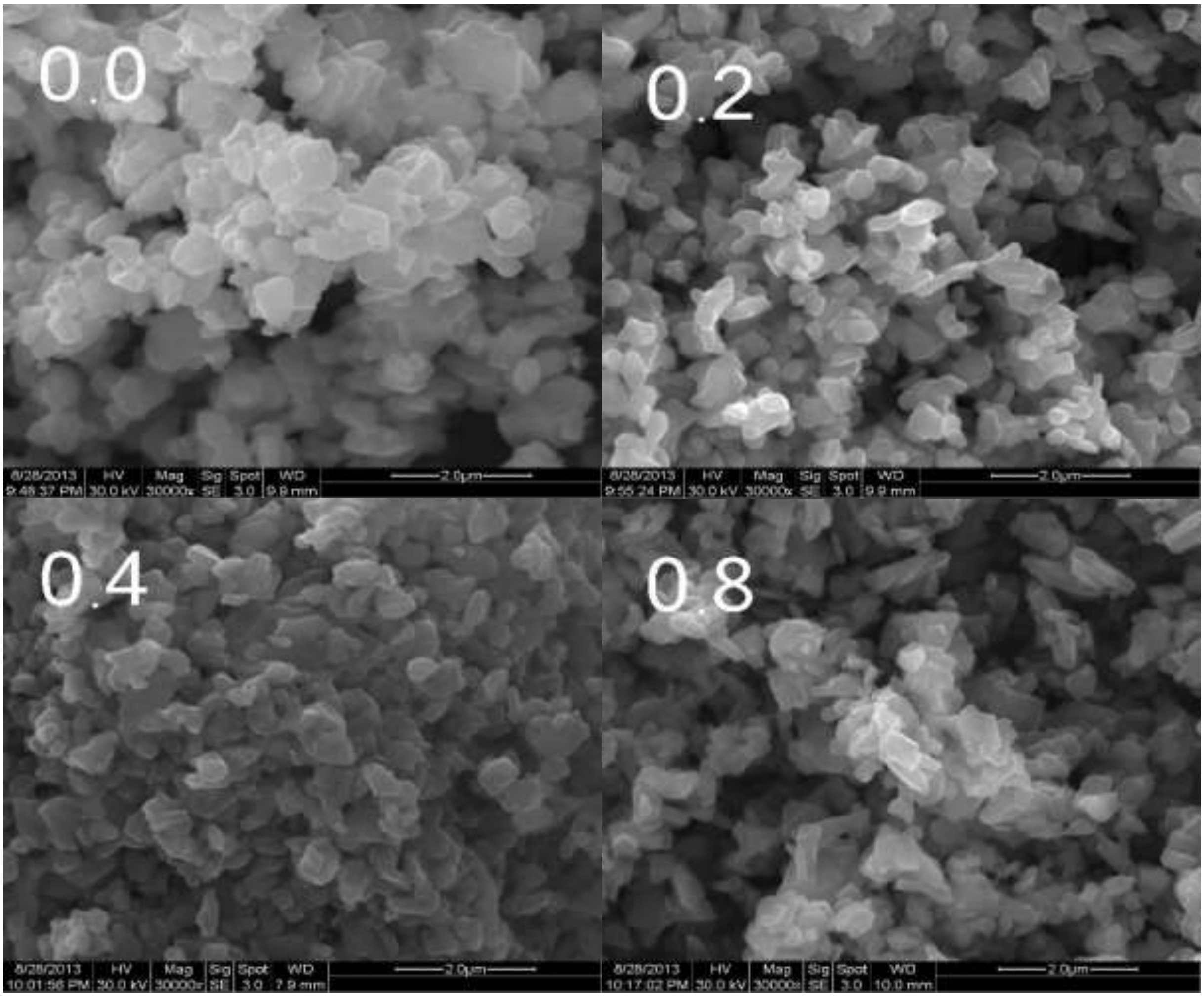

Fig. 3. Representative SEM images of $SrFe_{12-2x}Co_xTi_xO_{19}$ ($x$ = 0.0, 0.2, 0.4, 0.8).

### 4.4 MAGNETISM

Fig. 4 shows the measured hysteresis loops for representative $SrFe_{12-2x}Co_xTi_xO_{19}$ samples as a function of applied magnetic field. The magnetization curve for the non-substituted sample belongs to a hard magnetic material with high intrinsic coercivity of 4400 Oe. This value is in agreement with the previously reported values for SrM hexaferrites [115, 165]. The magnetization of all samples did not saturate at the highest applied field, and the law of approach to saturation [63, 166, 167] was employed to obtain the saturation magnetization of the samples. The effect of Co–Ti substitution on the saturation magnetization and coercivity of $SrFe_{12-2x}Co_xTi_xO_{19}$ are shown in Fig. 5. With increasing *x* the saturation magnetization starts increasing for *x* > 0.6, recording an increase of 10% for the sample *x* = 0.8. Fig 5 also shows that the intrinsic coercivity drops dramatically from about 4400 Oe to 700 Oe for *x* = 1.0. The remanent magnetization decreased slightly from 39.5 emu/g for the sample with *x* = 0.2 down to 36.6 emu/g for the sample with *x* = 0.8, while the coercivity dropped down to 1425 Oe for the latter sample. Thus, the substitution of Fe by Co–Ti in SrM ferrite with *x* up to 0.8 resulted in a substantial reduction in coercivity, and maintained the saturation magnetization and remanence at relatively high levels suitable for high density magnetic recording applications.

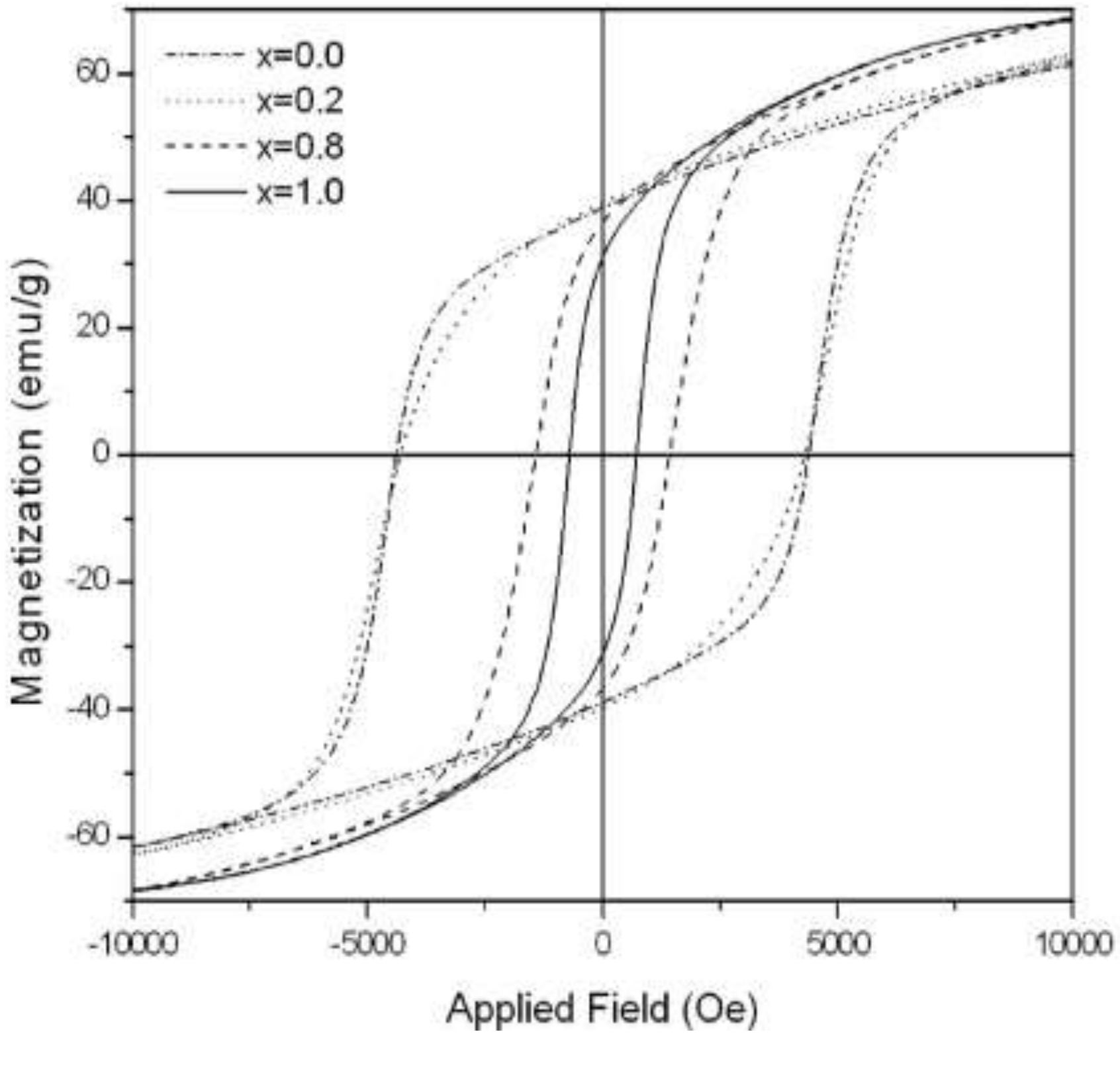

Fig. 4. Hysteresis loops for some of the $SrFe_{12-2x}Co_xTi_xO_{19}$ (*x* = 0.0, 0.2, 0.8, 1.0) samples as a function of applied magnetic field.

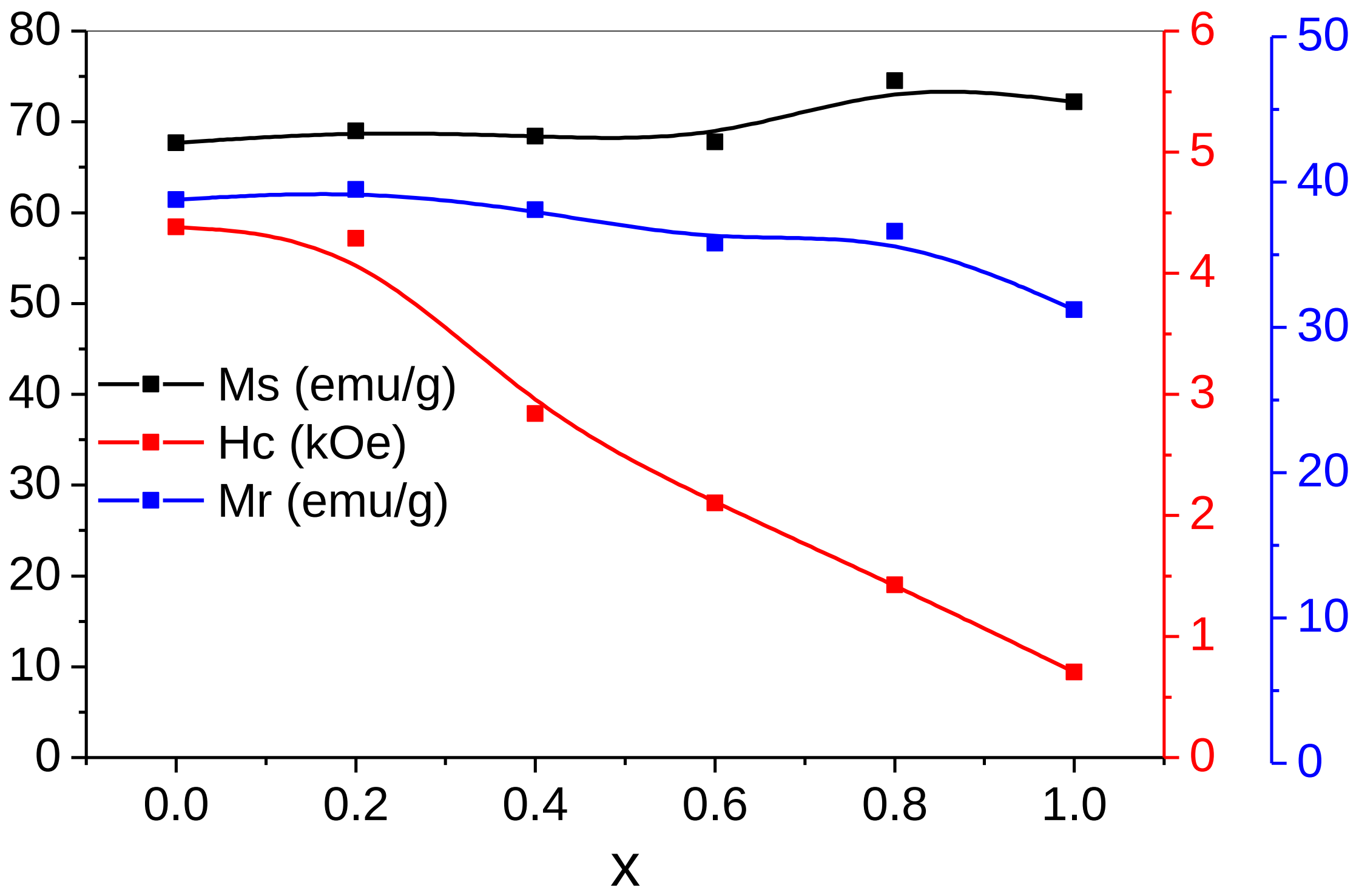

Fig. 5. Saturation magnetization, remanent magnetization, and intrinsic coercivity of $SrFe_{12-2x}Co_xTi_xO_{19}$ as a function of the concentration ($x$).

Ferric ions in M-type hexaferrites occupy five different interstitial sites, three of which are octahedral (12$k$, 4$f_2$ and 2$a$), one is tetrahedral (4$f_1$) and one is trigonal bi-pyramidal (2$b$). The crystal symmetry involving Fe–O bond lengths and Fe–O–Fe bond angles between the magnetic ions in the hexaferrite lattice lead to superexchange interactions which split the magnetic structure of M-type hexaferrite into spin-up and spin-down sublattices. These sublattices and their major contributions to the magnetic properties are summarized in Table 3.

*Table 3. Magnetic sublattices and their contribution to the magnetic properties of M-type hexaferrites.*

| Sites | Coordination | Spin direction | Contribution |
|---|---|---|---|
| 4$f_2$ | Octahedral | Down | Magnetization; Coercivity; Anisotropy |
| 2$b$ | Bi-pyramid | Up | Magnetization; Coercivity; Anisotropy |
| 12$k$ | Octahedral | Up | Magnetization; Anisotropy |
| 2$a$ | Octahedral | Up | Magnetization |
| 4$f_1$ | Tetrahedral | Down | Magnetization |

It was reported that the major positive contribution to the magnetic anisotropy in M-type hexaferrites comes from iron ions at $4f_2$ and $2b$ sites, while the contribution of $12k$ site is negative, preferring in-plane easy-axis [168]. Neutron diffraction results on Co–Ti substituted BaM indicated that 50% of the $Co^{2+}$ ions occupy $4f_1$ site, and $Ti^{4+}$ ions preferentially $4f_2$ site [144]. However, evidence of substitution at $12k$ and $2b$ sites was provided [168]. On the other hand, analysis of the site preference based on Mössbauer spectroscopy results indicated that the substitution occurs at $4f_2$ and $2b$ sites, and that the $12k$ is not a preferred substitutional site [145]. Further, Mössbauer spectra of Co–Sn substituted BaM indicated preference for substitution at $4f_2$, $2b$, and $12k$ sites [123]. The saturation magnetization did not change significantly with increasing $x$, up to 0.6, indicating that in this range of substitution, Co and Ti ions are distributed almost equally among spin-up and spin-down sublattices. For higher values of $x$, the slight increase in saturation magnetization could be ascribed to the involvement of the $4f_1$ spin-down sublattice as a substitutional site at high Co–Ti concentrations [145, 148, 168].

The dramatic drop of the intrinsic coercivity (Fig. 5) could not be ascribed to changes in grain size and morphology, since Co–Ti did not induce noticeable microstructural effects. This behavior is therefore associated with intrinsic parameters involving the reduction of the magnetocrystalline anisotropy. In order to investigate the effect of Co–Ti substitution on the magnetic anisotropy, the effective anisotropy field ($H_a$) was determined for each sample examined in this work from the switching field distribution (SFD). The switching field distribution can be obtained by differentiating the reduced IRM curve $m_r(H) = M_r(H)/M_r(\infty)$. The effective magnetic anisotropy field for each sample examined in this work is obtained from the maximum of the switching field distribution according to the formula [169]:

$$f(H)\big]_{\max} = \left[\frac{dm_r}{dH}\right]_{H=H_a/2} \qquad (2)$$

Here $H_a = 2H_{max}$, where $H_{max}$ is the value of the field at the maximum of the SFD. Fig. 6 shows representative reduced IRM curves and the corresponding switching field distributions for the samples with $x = 0.0$ and 1.0.

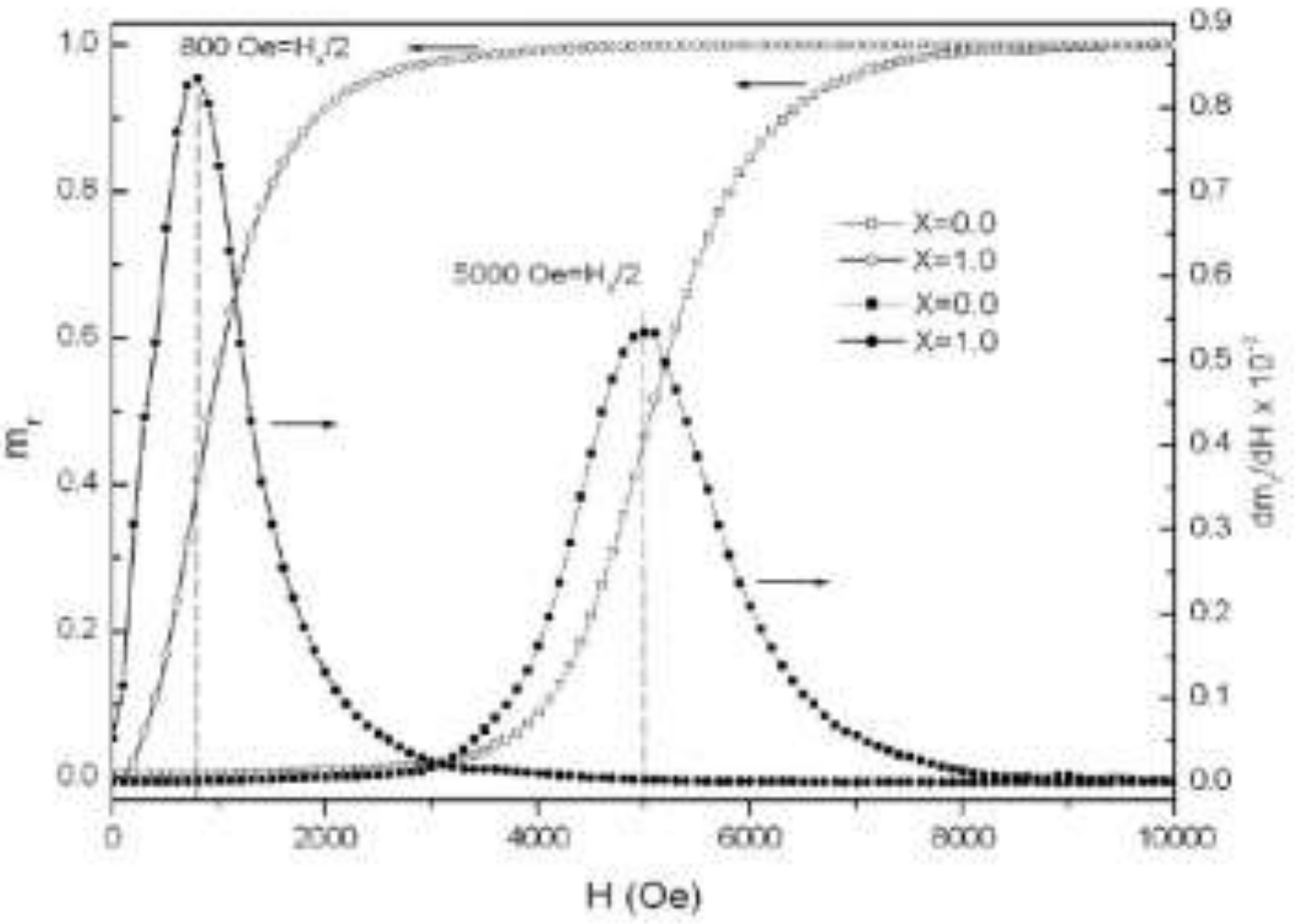

Fig. 6. Reduced IRM curves and switching field distributions for the samples with *x* = 0.0 and *x* = 1.0.

Fig. 7 shows the variation of magnetic anisotropy field with Co-Ti concentration for all samples examined. It is clear that $H_a$ decreases monotonically with increasing Co-Ti concentration beyond $x = 0.2$. Considering the results of SEM imaging which revealed similarity of the particle shape and size distribution in these samples, the monotonic decline in the intrinsic coercivity for $x \geq 0.2$ is well explained by the dependence of the anisotropy field on Co–Ti concentration. This result is consistent with the reported decrease of the anisotropy constant with increasing Co–Ti, and the consequent decrease in coercivity [147].

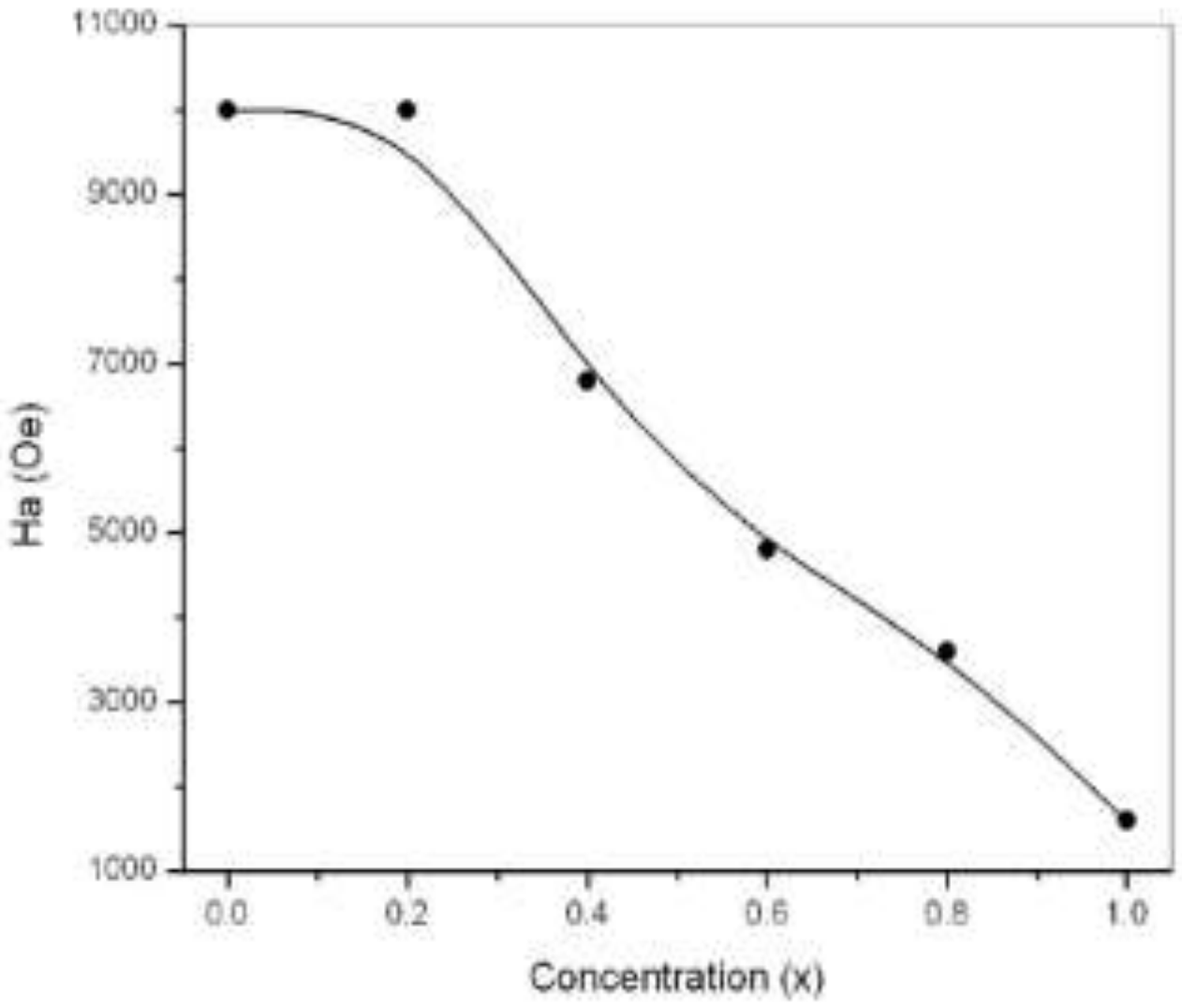

Fig. 7. Anisotropy field of $SrFe_{12-2x}Co_xTi_xO_{19}$ as a function of the Co–Ti concentration.

The *B-H* and *J-H* curves of the samples under investigation are shown in Fig. 8, where $J = 4\pi M$, $M$ being the magnetization in emu/cm$^3$. From these curves, the Retentivity (or residual induction $B_r$), the coercivity $H_{cB}$, and the intrinsic coercivity $H_{cJ}$ can be determined. These magnetic parameters are important for evaluating the suitability of a magnetic material for a given application. Magnetic materials with high residual induction and high coercivity are sought for permanent magnet industry. For magnetic recording applications, however, the coercivity should be reduced to ≳ 1000 Oe, while maintaining the residual induction as high as possible to provide good signal and avoid erasure of stored information by stray fields [121, 122]. The curves in Fig. 8 indicate that the sample with $x$ = 0 is characterized by a linear *B-H* curve far into the second quadrant of the hysteresis loop, indicating that this sample is suitable as a permanent magnet with relatively high coercivity. The *B-H* curves deviate progressively from linearity, and the residual induction decreases with increasing Co–Ti concentration as demonstrated by the induction curves in the second quadrant (Fig. 9).

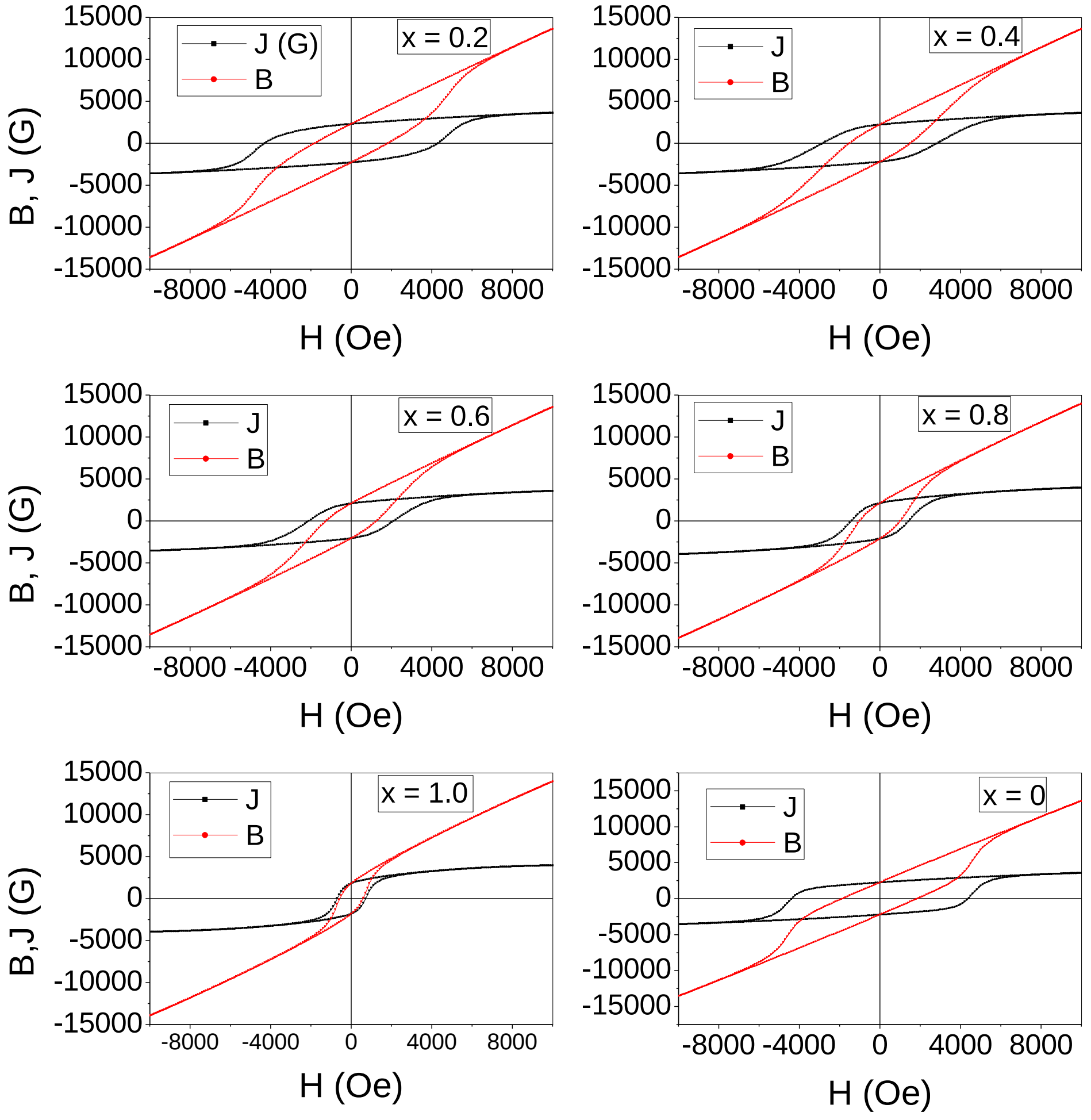

Fig. 8. *B-H* and *J-H* curves for Co–Ti substituted SrM (($x$ = 0.0, 0.2, 0.4, 0.6, 0.8, 1.0).

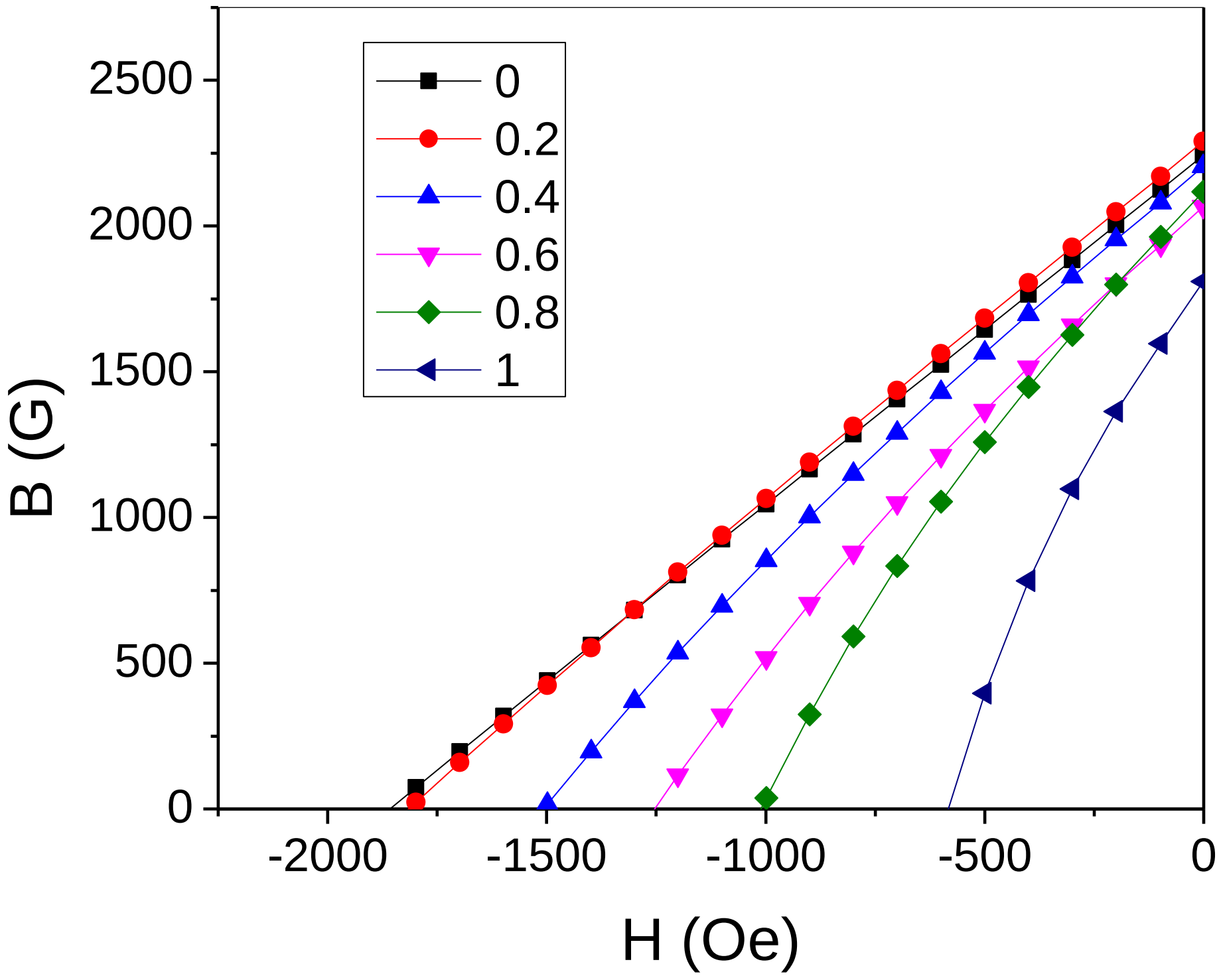

Fig. 9. *B-H* curves in the second quadrant for Co–Ti substituted SrM.

The maximum energy product $(BH)_{max}$ is a quality factor for a permanent magnet. A plot of *BH* vs. –*H* in the second quadrant of the hysteresis loop for the sample with $x = 0$ is shown in Fig. 10. The curve indicates that this sample is characterized by $(BH)_{max} =$ 1.05 MGOe (8.35 kJ/m$^3$). Although this value is significantly smaller than that for RE-permanent magnets, SrM remains of practical importance for permanent magnet industry, considering the low cost of production, availability of raw materials, and corrosion resistance. The maximum energy product, as well as the remanent induction and coercivity for all samples are evaluated from the *B-H* and *J-H* curves, and the results are summarized in Table 4. The magnetic properties of the samples indicate that these hexaferrites demonstrate potential for permanent magnet applications in the low *x* region (up to 0.2), whereas at higher substitution levels, the ferrites are suitable for high density magnetic recording applications.

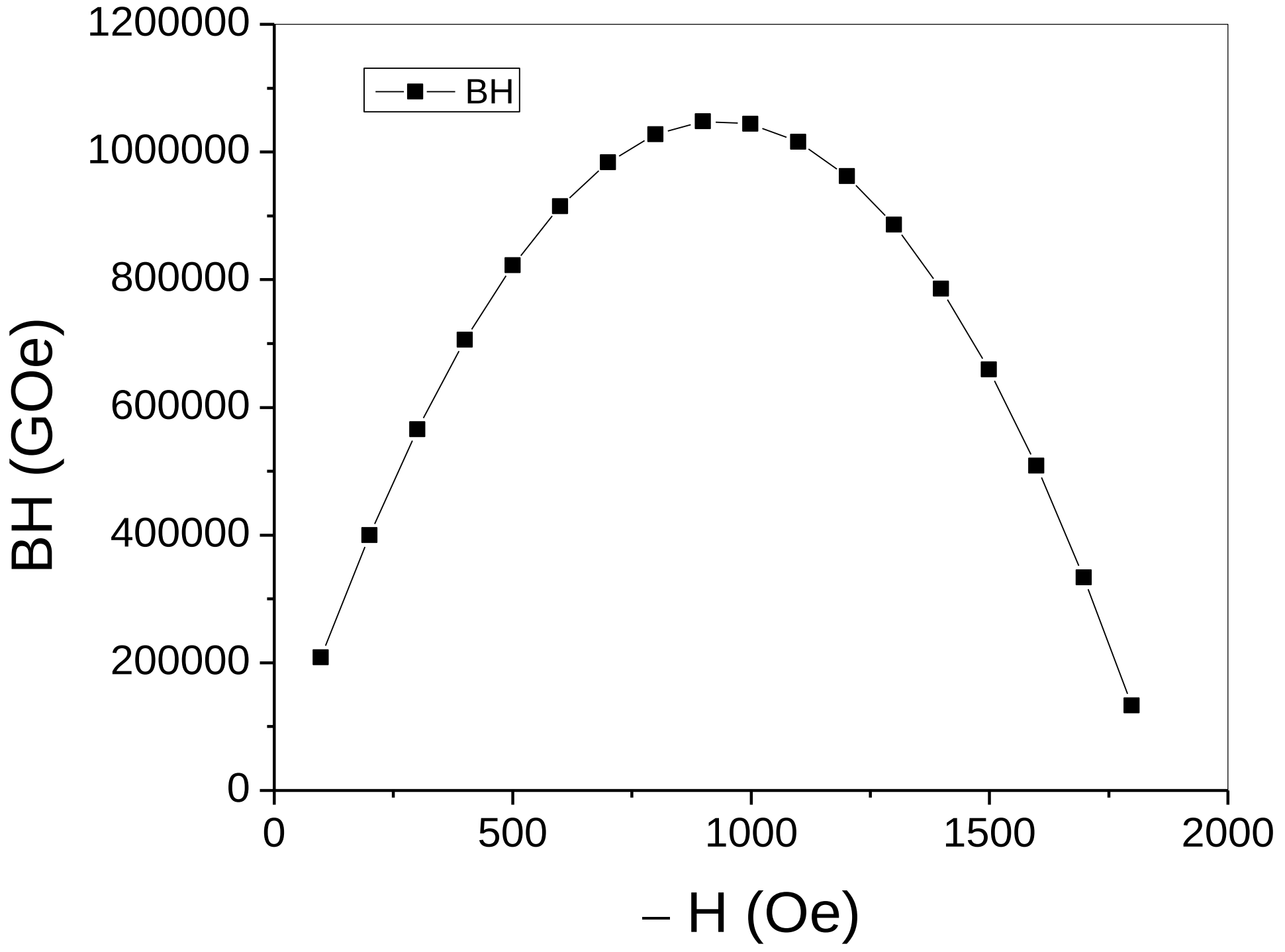

Fig. 10. *BH* curve in the second quadrant of the hysteresis loop the pure SrM sample.

*Table 4. Magnetic parameters of Co-Ti substituted SrM ferrites.*

| $x$ | $B_r$ (G) | $H_{cB}$(Oe) | $H_{cJ}$ (Oe) | $(BH)_{max}$ (kJ/m$^3$) |
|---|---|---|---|---|
| 0.0 | 2240 | 1850 | 4400 | 8.35 |
| 0.2 | 2288 | 1815 | 4300 | 8.52 |
| 0.4 | 2204 | 1500 | 2840 | 7.32 |
| 0.6 | 2066 | 1250 | 2200 | 5.84 |
| 0.8 | 2110 | 1025 | 1425 | 5.02 |
| 1.0 | 1800 | 575 | 620 | 2.62 |

The temperature dependence of the specific magnetization M(T) was investigated (Fig.11). The magnetization was measured as a function of temperature in a constant applied field of 100 Oe. All curves exhibited Hopkinson peak just below Curie temperature ($T_c$), which is characteristic of the presence of superparamagnetic grains in the samples [170]. From these curves, it is clear that the fraction of the superparamagnetic phases is reduced with Co–Ti substitution. The broadening and induced irregularity in the peak structure at high substitution levels could be due to magnetic inhomogeneity of the sample. Curie temperature as a function of Co-Ti concentration in $SrFe_{12-2x}Co_xTi_xO_{19}$ is shown in Table 5. The decrease in $T_c$ is consistent with the conclusion that the replacement of Fe ions by Co-Ti ions results in the attenuation of the magnetic exchange coupling.

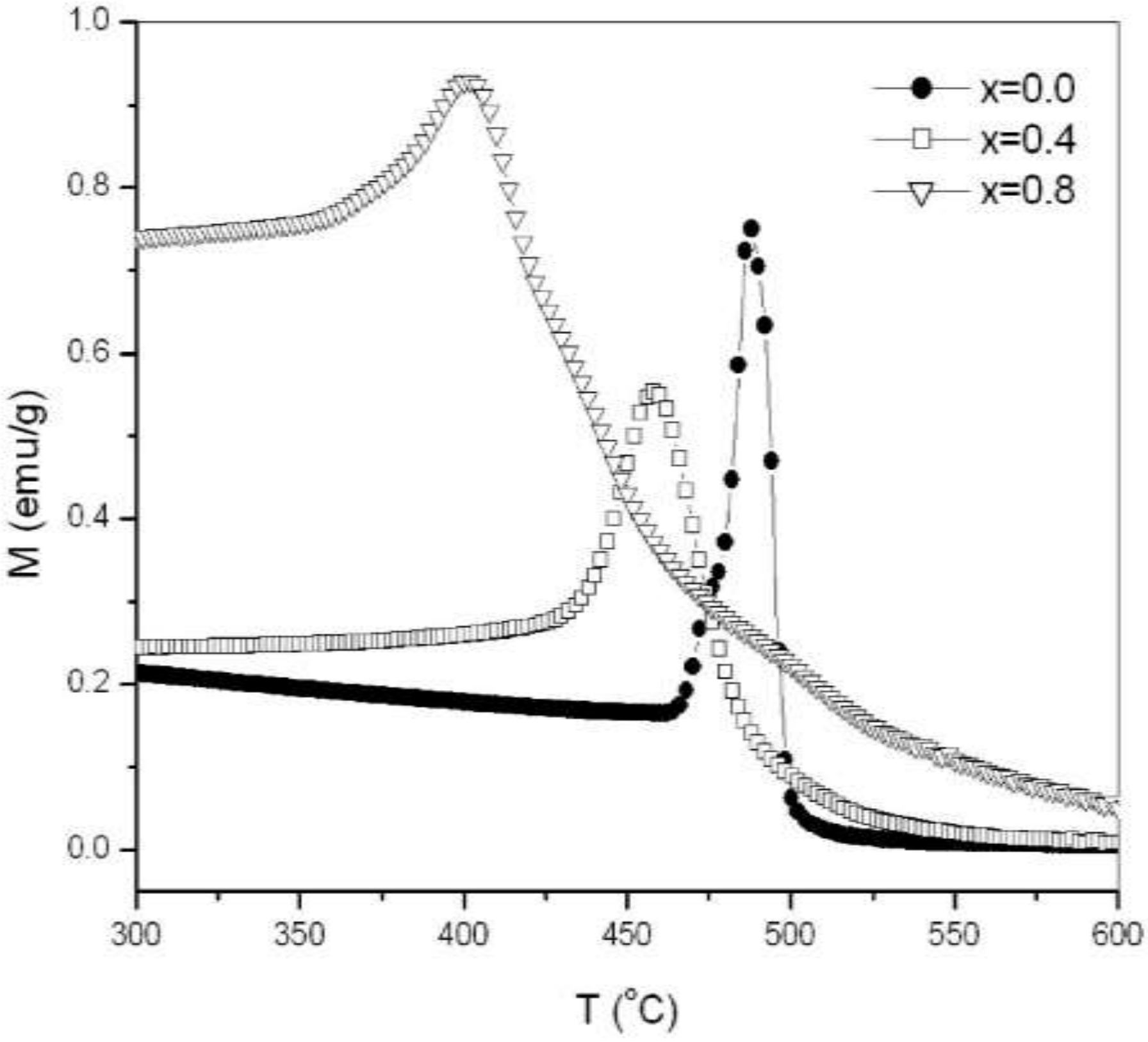

Fig. 11. Magnetization as a function of temperature for $SrFe_{12-2x}Co_xTi_xO_{19}$ with $x$ = 0.0, 0.4 and 0.8.

*Table 5. Curie temperature for $SrFe_{12-2x}Co_xTi_xO_{19}$.*

| $x$ | 0.0 | 0.2 | 0.4 | 0.6 | 0.8 | 1.0 |
|---|---|---|---|---|---|---|
| $T_c$ (°C) | 498 | 490 | 478 | 430 | 420 | 368 |

## 4.5 INTER-PARTICLE INTERACTIONS

In order to investigate the role of doping with Co-Ti on the inter-particle interactions in the prepared SrM ferrites, we evaluate $\Delta m$ defined by the relation [171]:

$$\Delta m(H) = m_d(H) - \left[1 - 2m_r(H)\right] \qquad (3)$$

where:

$$m_r(H) = M_r(H)/M_r(\infty) \qquad (4)$$

is the reduced isothermal remanent magnetization (IRM), and

$$m_d(H) = M_d(H)/M_r(\infty) \qquad (5)$$

is the reduced dc demagnetization (DCD). The IRM curve was obtained by the following procedure: the sample was fist demagnetized, then a positive field is applied, and finally, the remanent magnetization is measured after removing the applied field. The procedure was repeated with increasing the positive field to reach positive saturation remanence. The DCD curve was obtained by first saturating the sample with a positive field of 10 kOe, then a negative field was applied to the sample, and the remanent magnetization was recorded after removing the negative field; this procedure was repeated with increasing the negative field until negative saturation remanence was reached.

The IRM and DCD curves for all samples examined in this work are shown on Fig. 12. $\Delta m(H)$, which gives the strength and the sign of the inter-particle interactions was evaluated for each sample using equation 3 and the data in Fig. 12. For an assembly of non-interacting particles, $\Delta m$ is field-independent; any deviation from this behavior is a sign of the existence of inter-particle interactions. Positive $\Delta m$ values indicate the existence of inter-particle interactions that contribute constructively to the magnetization (magnetizing-like effect), where particles tend to stack in columns. On the other hand, negative $\Delta m$ values suggest that the existing interactions are demagnetizing (demagnetizing-like effect), where the particles tend to form clusters.

Fig. 13 shows the $\Delta m$ curves as a function of the applied field for samples with different concentrations of Co-Ti. It is clear that $\Delta m$ values for all samples are negative at all fields, with a maximum value occurring around the intrinsic coercivity. This result is an indication of a demagnetizing-like effect in all samples, in agreement with SEM observation of cluster formation. Also, the maximum negative value of $\Delta m$ decreased with increasing Co-Ti concentration, indicating that Co-Ti substitution may result in the suppression of cluster formation.

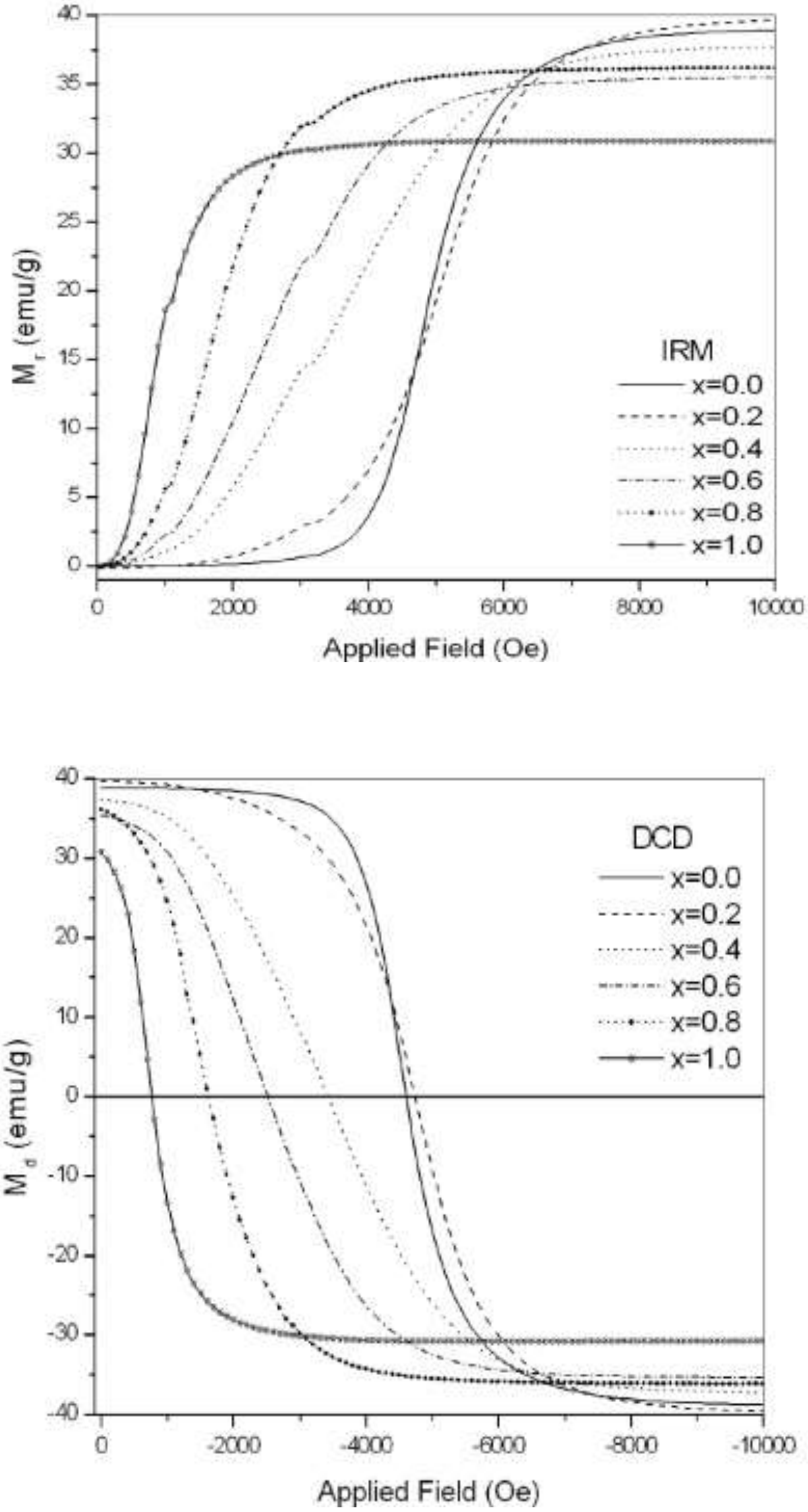

Fig. 12. IRM and DCD curves of $SrFe_{12-2x}Co_xTi_xO_{19}$.

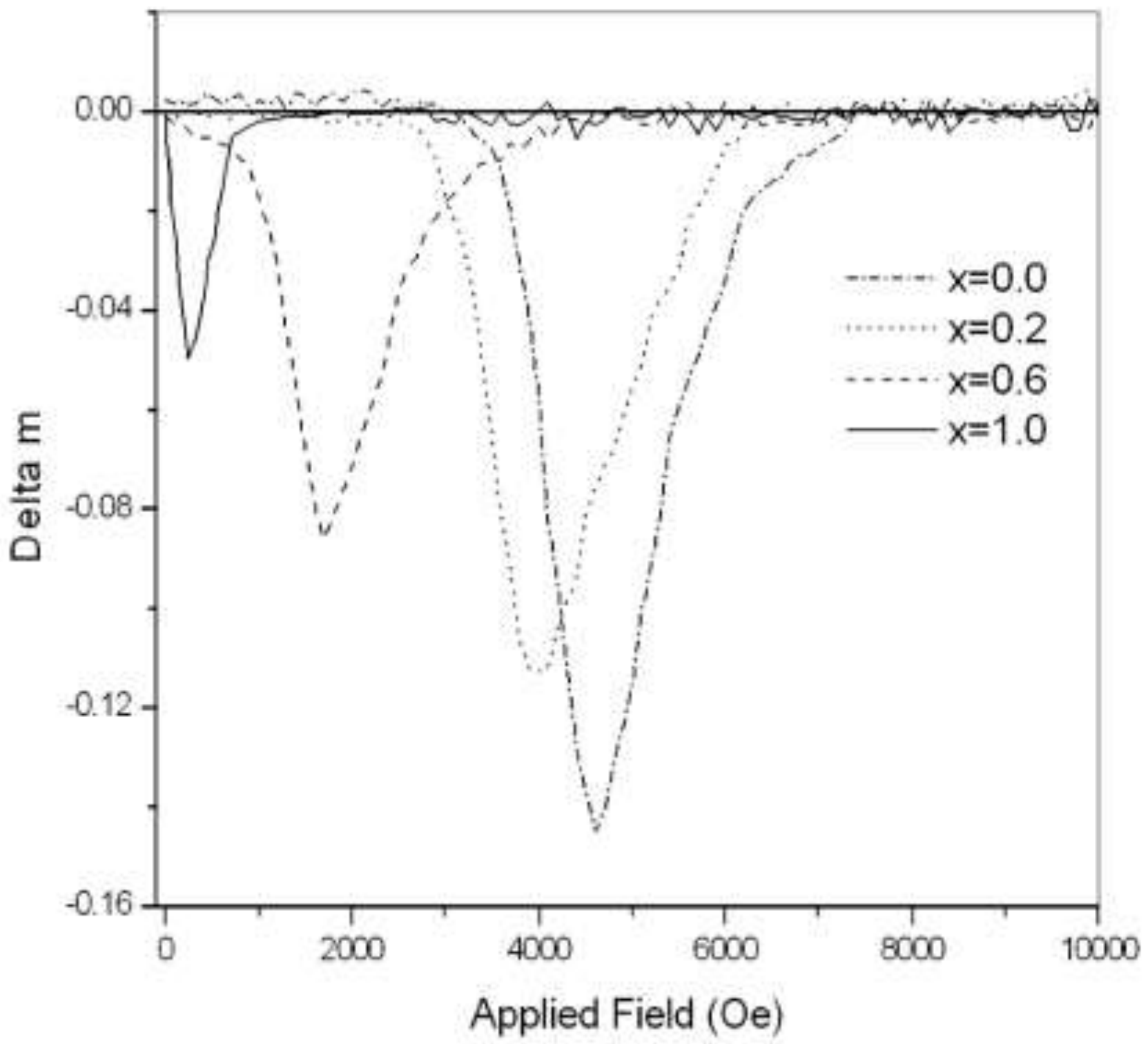

Fig. 13. Delta M curves of $SrFe_{12-2x}Co_xTi_xO_{19}$ for all concentration examined.

## 5. CONCLUSIONS

**The magnetic properties of hexaferrites in current use, or those with potential for future applications, span the wide range of coercivity from few hundred Oe to several thousands of Oe, depending on the application to be used for. In this article, the enhancement of the magnetic properties of M-type hexaferrites, and modifications made by adopting different synthesis routes and experimental conditions are reviewed. Chemical methods were found to be more effective than the conventional solid state reaction in controlling the particle shapes and particle size distribution, and improving the magnetic properties. Complex experimental procedures including variations of the chemical stoichiometry, heat treatment, and combinations of different synthesis techniques were adopted for further improvements of the magnetic properties of the ferrite powders. Also, the effects of the various types of substitutions on the magnetic properties of the hexaferrites were reviewed. Since the term "improvement of the properties" is device designer dependent, substitutions leading to the improvement of the magnetic properties of the ferrites for different types of device applications were addressed. Generally, the substitution of Fe ions by trivalent ions such as Al or Cr ions results in a significant increase in the coercivity suitable for permanent magnet applications. The coercivity was further enhanced by rare-earth metal substitutions. Also, the substitution of $Fe^{3+}$ ions in M-type hexaferrites by**

**combinations of divalent–tetravalent ions results in appreciable modifications of the magnetic properties the hexaferrites. In particular, Co-Ti substitution was found to be effective in modifying the properties of the ferrites to be suitable for high density magnetic recording and microwave applications. In addition, the structural and magnetic properties a series of $SrFe_{12-2x}Co_xTi_xO_{19}$ ferrites prepared by ball milling method were carefully investigated. The effect of Co–Ti substitution on the magnetic properties was found to reduce the coercivity down to levels suitable for high density longitudinal and perpendicular magnetic recording, without reducing the saturation magnetization. Although no further improvements concerning the magnetic properties of those magnetic materials are required for such applications, control over the particle size and shape, as well as the orientation of the magnetic particles on the substrate of the recording media may require special procedures.**

## REFERENCES

[1] http://www.magneticsmagazine.com/main/news/permanent-magnet-market-will-reach-28-70-billion-in-2019/.

[2] R.C. Pullar, Hexagonal ferrites: a review of the synthesis, properties and applications of hexaferrite ceramics, Progress in Materials Science, 57 (2012) 1191-1334.

[3] Ü. Özgür, Y. Alivov, H. Morkoç, Microwave ferrites, part 1: fundamental properties, Journal of Materials Science: Materials in Electronics, 20 (2009) 789-834.

[4] K.J. Strnat, Modern permanent magnets for applications in electro-technology, Proceedings of the IEEE, 78 (1990) 923-946.

[5] V.G. Harris, A. Geiler, Y. Chen, S.D. Yoon, M. Wu, A. Yang, Z. Chen, P. He, P.V. Parimi, X. Zuo, Recent advances in processing and applications of microwave ferrites, Journal of Magnetism and Magnetic Materials, 321 (2009) 2035-2047.

[6] J. Smit, H.P.J. Wijn, Ferrites, Wiley, New York, 1959.

[7] S. Chikazumi, Physics of Ferromagnetism 2e, Oxford University Press 2009.

[8] S.H. Mahmood, A.N. Aloqaily, Y. Maswadeh, A. Awadallah, I. Bsoul, M. Awawdeh, H.K. Juwhari, Effects of heat treatment on the phase evolution, structural, and magnetic properties of Mo-Zn doped M-type hexaferrites, Solid State Phenomena, 232 (2015) 65-92.

[9] S.H. Mahmood, M.D. Zaqsaw, O.E. Mohsen, A. Awadallah, I. Bsoul, M. Awawdeh, Q.I. Mohaidat, Modification of the magnetic properties of $Co_2Y$ hexaferrites by divalent and trivalent metal substitutions, Solid State Phenomena, 241 (2016) 93-125.

[10] G. Albanese, Recent advances in hexagonal ferrites by the use of nuclear spectroscopic methods, Le Journal de Physique Colloques, 38 (1977) 85-94.

[11] R. Chantrell, K. O'Grady, Magnetic characterization of recording media, Journal of Physics D: Applied Physics, 25 (1992) 1.

[12] H. Fu, H.R. Zhai, H.C. Zhang, B.X. Gu, J.Y. Li, Magnetic properties on Mn substituted barium ferrite, Journal of Magnetism and Magnetic Materials, 54-57 (1986) 905-906.

[13] I.Y. Gershov, Barium ferrite permanent magnets, Soviet Powder Metallurgy and Metal Ceramics, 1 (1964) 386-393.

[14] D. Lisjak, M. Drofenik, Synthesis and characterization of A–Sn-substituted (A= Zn, Ni, Co) BaM–hexaferrite powders and ceramics, Journal of the European Ceramic Society, 24 (2004) 1841-1845.

[15] S. Mahmood, A. Aloqaily, Y. Maswadeh, A. Awadallah, I. Bsoul, H. Juwhari, Structural and Magnetic Properties of Mo-Zn Substituted ($BaFe_{12-4x}Mo_xZn_{3x}O_{19}$) M-Type Hexaferrites, Material Science Research India, 11 (2014) 09-20.

[16] G. Turilli, F. Licci, S. Rinaldi, A. Deriu, $Mn^{2+}$, $Ti^{4+}$ substituted barium ferrite, Journal of Magnetism and Magnetic Materials, 59 (1986) 127-131.

[17] A. Awadallah, S.H. Mahmood, Y. Maswadeh, I. Bsoul, M. Awawdeh, Q.I. Mohaidat, H. Juwhari, Structural, magnetic, and Mossbauer spectroscopy of Cu substituted M-type hexaferrites, Materials Research Bulletin, 74 (2016) 192-201.

[18] O.T. Ozkan, H. Erkalfa, The effect of $B_2O_3$ addition on the direct sintering of barium hexaferrite, Journal of the European Ceramic Society, 14 (1994) 351-358.

[19] P. Hernandez-Gomez, J.M. Munoz, C. Torres, C. de Francisco, O. Alejos, Influence of stoichiometry on the magnetic disaccommodation in barium M-type hexaferrites, Journal of Physics D: Applied Physics, 36 (2003) 1062-1070.

[20] Y. Maswadeh, S.H. Mahmood, A. Awadallah, A.N. Aloqaily, Synthesis and structural characterization of nonstoichiometric barium hexaferrite materials with Fe: Ba ratio of 11.5–16.16, IOP Conference Series: Materials Science and Engineering, IOP Publishing, 2015, pp. 012019.

[21] Y.-M. Kang, Y.-H. Kwon, M.-H. Kim, D.-Y. Lee, Enhancement of magnetic properties in Mn–Zn substituted M-type Sr-hexaferrites, Journal of Magnetism and Magnetic Materials, 382 (2015) 10-14.

[22] Y. Maswadeh, Structural analysis of hexaferrite materials, Physics, The University of Jordan, 2014.

[23] P. Garcia-Casillas, A. Beesley, D. Bueno, J. Matutes-Aquino, C. Martinez, Remanence properties of barium hexaferrite, Journal of alloys and compounds, 369 (2004) 185-189.

[24] D. Lisjak, M. Drofenik, The mechanism of the low-temperature formation of barium hexaferrite, Journal of the European Ceramic Society, 27 (2007) 4515-4520.

[25] J.-P. Wang, L. Ying, M.-L. Zhang, Y.-j. QIAO, X. Tian, Comparison of the Sol-gel Method with the Coprecipitation Technique for Preparation of Hexagonal Barium Ferrite, Chemical Research in Chinese Universities, 24 (2008) 525-528.

[26] V. Harikrishnan, P. Saravanan, R.E. Vizhi, D.R. Babu, V. Vinod, P. Kejzlar, M. Černík, Effect of annealing temperature on the structural and magnetic properties of CTAB-capped $SrFe_{12}O_{19}$ platelets, Journal of Magnetism and Magnetic Materials, 401 (2016) 775-783.

[27] H.B. von Basel, K.A. Hempel, Static magnetic properties of pressure-sintered barium ferrite, Journal of Magnetism and Magnetic Materials, 38 (1983) 316-318.

[28] S.E. Jacobo, C. Domingo-Pascual, R. Rodrigez-Clemente, M.A. Blesa, Synthesis of ultrafine particles of barium ferrite by chemical coprecipitation, Journal of Materials Science, 33 (1997) 1025-1028.

[29] M. Rashad, I. Ibrahim, A novel approach for synthesis of M-type hexaferrites nanopowders via the co-precipitation method, Journal of Materials Science: Materials in Electronics, 22 (2011) 1796-1803.

[30] A. Davoodi, B. Hashemi, Investigation of the effective parameters on the synthesis of strontium hexaferrite nanoparticles by chemical coprecipitation method, Journal of Alloys and Compounds, 512 (2012) 179-184.

[31] S.R. Janasi, D. Rodrigues, F.J. Landgraf, M. Emura, Magnetic properties of coprecipitated barium ferrite powders as a function of synthesis conditions, Magnetics, IEEE Transactions on, 36 (2000) 3327-3329.

[32] J. Matutes-Aquino, S. Dıaz-Castanón, M. Mirabal-Garcıa, S. Palomares-Sánchez, Synthesis by coprecipitation and study of barium hexaferrite powders, Scripta materialia, 42 (2000) 295-299.

[33] P. Shepherd, K.K. Mallick, R.J. Green, Magnetic and structural properties of M-type barium hexaferrite prepared by co-precipitation, Journal of magnetism and magnetic materials, 311 (2007) 683-692.

[34] Z. Mosleh, P. Kameli, A. Poorbaferani, M. Ranjbar, H. Salamati, Structural, magnetic and microwave absorption properties of Ce-doped barium hexaferrite, Journal of Magnetism and Magnetic Materials, 397 (2016) 101-107.

[35] M. Jamalian, An investigation of structural, magnetic and microwave properties of strontium hexaferrite nanoparticles prepared by a sol–gel process with doping SN and Tb, Journal of Magnetism and Magnetic Materials, 378 (2015) 217-220.

[36] W. Zhong, W. Ding, N. Zhang, J. Hong, Q. Yan, Y. Du, Key step in synthesis of ultrafine $BaFe_{12}O_{19}$ by sol-gel technique, Journal of Magnetism and Magnetic Materials, 168 (1997) 196-202.

[37] R.C. Alange, P.P. Khirade, S.D. Birajdar, A.V. Humbe, K.M. Jadhav, Structural, magnetic and dielectric properties of Al-Cr co-substituted M-type barium hexaferrite nanoparticles, Journal of Molecular Structure, 1106 (2016) 460-467.

[38] Y. Hong, C. Ho, H.Y. Hsu, C. Liu, Synthesis of nanocrystalline $Ba(MnTi)_xFe_{12-2x}O_{19}$ powders by the sol–gel combustion method in citrate acid–metal nitrates system ($x$ = 0, 0.5, 1.0, 1.5, 2.0), Journal of magnetism and magnetic materials, 279 (2004) 401-410.

[39] S.H. Mahmood, F.S. Jaradat, A.F. Lehlooh, A. Hammoudeh, Structural properties and hyperfine interactions in Co-Zn Y-type hexaferrites prepared by sol-gel method, Ceramics International, 40 (2014) 5231-5236.

[40] W. Abbas, I. Ahmad, M. Kanwal, G. Murtaza, I. Ali, M.A. Khan, M.N. Akhtar, M. Ahmad, Structural and magnetic behavior of Pr-substituted M-type hexagonal ferrites synthesized by sol–gel autocombustion for a variety of applications, Journal of Magnetism and Magnetic Materials, 374 (2015) 187-191.

[41] Simon Thompson, Neil J. Shirtcliffe, Eoin S. O'Keefe, Steve Appleton, C.C. Perry, Synthesis of $SrCo_xTi^xFe_{(12-2x)}O_{19}$ through sol-gel auto-ignition and its characterisation, Journal of Magnetism and Magnetic Materials, 297 (2005) 100-1007.

[42] Y. Meng, M. He, Q. Zeng, D. Jiao, S. Shukla, R. Ramanujan, Z. Liu, Synthesis of barium ferrite ultrafine powders by a sol–gel combustion method using glycine gels, Journal of Alloys and Compounds, 583 (2014) 220-225.

[43] D. Bahadur, S. Rajakumar, A. Kumar, Influence of fuel ratios on auto combustion synthesis of barium ferrite nano particles, Journal of chemical sciences, 118 (2006) 15-21.

[44] V. Sankaranarayanan, Q. Pankhurst, D. Dickson, C. Johnson, Ultrafine particles of barium ferrite from a citrate precursor, Journal of magnetism and magnetic materials, 120 (1993) 73-75.

[45] V. Sankaranarayanan, D. Khan, Mechanism of the formation of nanoscale M-type barium hexaferrite in the citrate precursor method, Journal of magnetism and magnetic materials, 153 (1996) 337-346.

[46] V. Sankaranarayanan, Q. Pankhurst, D. Dickson, C. Johnson, An investigation of particle size effects in ultrafine barium ferrite, Journal of magnetism and magnetic materials, 125 (1993) 199-208.

[47] X. Liu, J. Wang, L.-M. Gan, S.-C. Ng, Improving the magnetic properties of hydrothermally synthesized barium ferrite, Journal of magnetism and magnetic materials, 195 (1999) 452-459.

[48] A. Ataie, I. Harris, C. Ponton, Magnetic properties of hydrothermally synthesized strontium hexaferrite as a function of synthesis conditions, Journal of materials science, 30 (1995) 1429-1433.

[49] D. Primc, D. Makovec, D. Lisjak, M. Drofenik, Hydrothermal synthesis of ultrafine barium hexaferrite nanoparticles and the preparation of their stable suspensions, Nanotechnology, 20 (2009) 315605.

[50] M. Drofenik, I. Ban, D. Makovec, A. Žnidaršič, Z. Jagličić, D. Hanžel, D. Lisjak, The hydrothermal synthesis of super-paramagnetic barium hexaferrite particles, Materials Chemistry and Physics, 127 (2011) 415-419.

[51] R.H. Arendt, The molten salt synthesis of single domain $BaFe_{12}O_{19}$ and $SrFe_{12}O_{19}$ crystals, Journal of Solid State Chemistry, 8 (1973) 339-347.

[52] T.-S. Chin, S. Hsu, M. Deng, Barium ferrite particulates prepared by a salt-melt method, Journal of magnetism and magnetic materials, 120 (1993) 64-68.

[53] Y. Liu, M.G. Drew, Y. Liu, J. Wang, M. Zhang, Preparation, characterization and magnetic properties of the doped barium hexaferrites $BaFe_{12-2x}Co_{x/2}Zn_{x/2}Sn_xO_{19}$, $x$ = 0.0–2.0, Journal of Magnetism and Magnetic Materials, 322 (2010) 814-818.

[54] O. Kubo, T. Ido, H. Yokoyama, Properties of Ba ferrite particles for perpendicular magnetic recording media, Magnetics, IEEE Transactions on, 18 (1982) 1122-1124.

[55] O. Kubo, T. Ido, T. Nomura, K. Inomata, Method for manufacturing magnetic powder for high density magnetic recording, Google Patents, 1982.

[56] B. Shirk, W. Buessem, Magnetic properties of barium ferrite formed by crystallization of a glass, Journal of the American Ceramic Society, 53 (1970) 192-196.

[57] D. Jung, S. Hong, J. Cho, Y. Kang, Nano-sized barium titanate powders with tetragonal crystal structure prepared by flame spray pyrolysis, Journal of the European Ceramic Society, 28 (2008) 109-115.

[58] J.S. Cho, D.S. Jung, S.K. Hong, Y.C. Kang, Characteristics of nano-sized pb-based glass powders by high temperature spray pyrolysis method, Journal of the Ceramic Society of Japan, 116 (2008) 600-604.

[59] D.-H. Kim, Y.-K. Lee, K.-M. Kim, K.-N. Kim, S.-Y. Choi, I.-B. Shim, Synthesis of Ba-ferrite microspheres doped with Sr for thermoseeds in hyperthermia, Journal of materials science, 39 (2004) 6847-6850.

[60] M.H. Kim, D.S. Jung, Y.C. Kang, J.H. Choi, Nanosized barium ferrite powders prepared by spray pyrolysis from citric acid solution, Ceramics International, 35 (2009) 1933-1937.

[61] U. Topal, Improvement of the remanence properties and the weakening of interparticle interactions in $BaFe_{12}O_{19}$ particles by $B_2O_3$ addition, Physica B: Condensed Matter, 407 (2012) 2058-2062.

[62] U. Topal, Towards Further Improvements of the Magnetization Parameters of $B_2O_3$-Doped $BaFe_{12}O_{19}$ Particles: Etching with Hydrochloric Acid, Journal of superconductivity and novel magnetism, 25 (2012) 1485-1488.

[63] A. Awadallah, S.H. Mahmood, Y. Maswadeh, I. Bsoul, A. Aloqaily, Structural and magnetic properties of Vanadium Doped M-Type Barium Hexaferrite ($BaFe_{12-x}V_xO_{19}$), IOP Conference Series: Materials Science and Engineering, IOP Publishing, 2015, pp. 012006.

[64] Q. Mohsen, Barium hexaferrite synthesis by oxalate precursor route, Journal of Alloys and Compounds, 500 (2010) 125-128.

[65] T. Gonzalez-Carreno, M. Morales, C. Serna, Barium ferrite nanoparticles prepared directly by aerosol pyrolysis, materials letters, 43 (2000) 97-101.

[66] U. Topal, H. Ozkan, H. Sozeri, Synthesis and characterization of nanocrystalline $BaFe_{12}O_{19}$ obtained at 850 C by using ammonium nitrate melt, Journal of magnetism and magnetic materials, 284 (2004) 416-422.

[67] U. Topal, H. Ozkan, L. Dorosinskii, Finding optimal Fe/Ba ratio to obtain single phase $BaFe_{12}O_{19}$ prepared by ammonium nitrate melt technique, Journal of alloys and compounds, 428 (2007) 17-21.

[68] S. El-Sayed, T. Meaz, M. Amer, H. El Shersaby, Magnetic behavior and dielectric properties of aluminum substituted M-type barium hexaferrite, Physica B: Condensed Matter, 426 (2013) 137-143.

[69] V.V. Soman, V. Nanoti, D. Kulkarni, Dielectric and magnetic properties of Mg–Ti substituted barium hexaferrite, Ceramics International, 39 (2013) 5713-5723.

[70] H. Sözeri, Effect of pelletization on magnetic properties of $BaFe_{12}O_{19}$, Journal of Alloys and Compounds, 486 (2009) 809-814.

[71] I. Bsoul, S. Mahmood, Structural and magnetic properties of $BaFe_{12-x}Al_xO_{19}$ prepared by milling and calcination, Jordan Journal of Physics, 2 (2009) 171-179.

[72] I. Bsoul, S. Mahmood, Magnetic and structural properties of $BaFe_{12-x}Ga_xO_{19}$ nanoparticles, Journal of Alloys and Compounds, 489 (2010) 110-114.

[73] I. Bsoul, S. Mahmood, A.-F. Lehlooh, Structural and magnetic properties of $BaFe_{12-2x}Ti_xRu_xO_{19}$, Journal of Alloys and Compounds, 498 (2010) 157-161.

[74] H.-F. Yu, $BaFe_{12}O_{19}$ powder with high magnetization prepared by acetone-aided coprecipitation, Journal of Magnetism and Magnetic Materials, 341 (2013) 79-85.

[75] Y. Liu, M.G. Drew, Y. Liu, Preparation and magnetic properties of barium ferrites substituted with manganese, cobalt, and tin, Journal of Magnetism and Magnetic Materials, 323 (2011) 945-953.

[76] E. Pashkova, E. Solovyova, I. Kotenko, T. Kolodiazhnyi, A. Belous, Effect of preparation conditions on fractal structure and phase transformations in the synthesis of nanoscale M-type barium hexaferrite, Journal of Magnetism and Magnetic Materials, 323 (2011) 2497-2503.

[77] G. Litsardakis, I. Manolakis, C. Serletis, K. Efthimiadis, High coercivity Gd-substituted Ba hexaferrites, prepared by chemical coprecipitation, Journal of Applied Physics, 103 (2008) 07E501.

[78] W. Roos, H. Haak, C. Voigt, K. Hempel, Microwave absorption and static magnetic properties of coprecipitated barium ferrite, Le Journal de Physique Colloques, 38 (1977) C1-35-C31-37.

[79] S. Kanagesan, M. Hashim, S. Jesurani, T. Kalaivani, I. Ismail, Influence of Zn–Nb on the Magnetic Properties of Barium Hexaferrite, Journal of Superconductivity and Novel Magnetism, 27 (2014) 811-815.

[80] T. Kaur, A. Srivastava, Effect of pH on Magnetic Properties of Doped Barium Hexaferrite, International Journal of Research in Mechanical Engineering & Technology, 3 (2013) 171-173.

[81] F. Khademi, A. Poorbafrani, P. Kameli, H. Salamati, Structural, magnetic and microwave properties of Eu-doped barium hexaferrite powders, Journal of superconductivity and novel magnetism, 25 (2012) 525-531.

[82] Y. Li, Q. Wang, H. Yang, Synthesis, characterization and magnetic properties on nanocrystalline $BaFe_{12}O_{19}$ ferrite, Current applied physics, 9 (2009) 1375-1380.

[83] C. Sürig, D. Bonnenberg, K. Hempel, P. Karduck, H. Klaar, C. Sauer, Effects of Variations in Stoichiometry on M-Type Hexaferrites, Le Journal de Physique IV, 7 (1997) C1-315-C311-316.

[84] V.C. Chavan, S.E. Shirsath, M.L. Mane, R.H. Kadam, S.S. More, Transformation of hexagonal to mixed spinel crystal structure and magnetic properties of $Co^{2+}$ substituted $BaFe_{12}O_{19}$, Journal of Magnetism and Magnetic Materials, 398 (2016) 32-37.

[85] R.K. Mudsainiyan, A.K. Jassal, M. Gupta, S.K. Chawla, Study on structural and magnetic properties of nanosized M-type Ba-hexaferrites synthesized by urea assisted citrate precursor route, Journal of Alloys and Compounds, 645 (2015) 421-428.

[86] H. Sözeri, Z. Durmuş, A. Baykal, E. Uysal, Preparation of high quality, single domain $BaFe_{12}O_{19}$ particles by the citrate sol–gel combustion route with an initial Fe/Ba molar ratio of 4, Materials Science and Engineering: B, 177 (2012) 949-955.

[87] V.N. Dhage, M. Mane, M. Babrekar, C. Kale, K. Jadhav, Influence of chromium substitution on structural and magnetic properties of $BaFe_{12}O_{19}$ powder prepared by sol–gel auto combustion method, Journal of Alloys and Compounds, 509 (2011) 4394-4398.

[88] M. Han, Y. Ou, W. Chen, L. Deng, Magnetic properties of Ba-M-type hexagonal ferrites prepared by the sol–gel method with and without polyethylene glycol added, Journal of alloys and compounds, 474 (2009) 185-189.

[89] T. Yamauchi, Y. Tsukahara, T. Sakata, H. Mori, T. Chikata, S. Katoh, Y. Wada, Barium ferrite powders prepared by microwave-induced hydrothermal reaction and magnetic property, Journal of Magnetism and Magnetic Materials, 321 (2009) 8-11.

[90] S. Dursun, R. Topkaya, N. Akdoğan, S. Alkoy, Comparison of the structural and magnetic properties of submicron barium hexaferrite powders prepared by molten salt and solid state calcination routes, Ceramics International, 38 (2012) 3801-3806.

[91] G. Albanese, A. Deriu, Magnetic properties of Al, Ga, Sc, In substituted barium ferrites: a comparative analysis, Ceramics International, 5 (1979) 3-10.

[92] M.H. Shams, A.S. Rozatian, M.H. Yousefi, J. Valíček, V. Šepelák, Effect of $Mg^{2+}$ and $Ti^{4+}$ dopants on the structural, magnetic and high-frequency ferromagnetic properties of barium hexaferrite, Journal of Magnetism and Magnetic Materials, 399 (2016) 10-18.

[93] R. Pullar, A. Bhattacharya, The magnetic properties of aligned M hexa-ferrite fibres, Journal of magnetism and magnetic materials, 300 (2006) 490-499.

[94] A. Alsmadi, I. Bsoul, S. Mahmood, G. Alnawashi, F. Al-Dweri, Y. Maswadeh, U. Welp, Magnetic study of M-type Ru-Ti doped strontium hexaferrite nanocrystalline particles, Journal of Alloys and Compounds, 648 (2015) 419-427.

[95] R. Palomino, A.B. Miró, F. Tenorio, F.S. De Jesús, C.C. Escobedo, S. Ammar, Sonochemical assisted synthesis of $SrFe_{12}O_{19}$ nanoparticles, Ultrasonics sonochemistry, 29 (2016) 470-475.

[96] A. Bolarín-Miró, F. Sánchez-De Jesús, C.A. Cortes-Escobedo, S. Diaz-De La Torre, R. Valenzuela, Synthesis of M-type $SrFe_{12}O_{19}$ by mechanosynthesis assisted by spark plasma sintering, Journal of Alloys and Compounds, 643 (2015) S226-S230.

[97] S. Singhal, T. Namgyal, J. Singh, K. Chandra, S. Bansal, A comparative study on the magnetic properties of $MFe_{12}O_{19}$ and $MAlFe_{11}O_{19}$ (M= Sr, Ba and Pb) hexaferrites with different morphologies, Ceramics International, 37 (2011) 1833-1837.

[98] A. Guerrero-Serrano, T. Pérez-Juache, M. Mirabal-García, J. Matutes-Aquino, S. Palomares-Sánchez, Effect of barium on the properties of lead hexaferrite, Journal of superconductivity and novel magnetism, 24 (2011) 2307-2312.

[99] M.N. Ashiq, R.B. Qureshi, M.A. Malana, M.F. Ehsan, Synthesis, structural, magnetic and dielectric properties of zirconium copper doped M-type calcium strontium hexaferrites, Journal of Alloys and Compounds, 617 (2014) 437-443.

[100] P. Popa, E. Rezlescu, C. Doroftei, N. Rezlescu, Influence of calcium on properties of strontium and barium ferrites for magnetic media prepared by combustion, J. Optoelectron. Adv. Mater, 7 (2005) 1553-1556.

[101] Ashima, S. Sanghi, A. Agarwal, Reetu, Rietveld refinement, electrical properties and magnetic characteristics of Ca–Sr substituted barium hexaferrites, Journal of Alloys and Compounds, 513 (2012) 436-444.

[102] X. Gao, Y. Du, X. Liu, P. Xu, X. Han, Synthesis and characterization of Co–Sn substituted barium ferrite particles by a reverse microemulsion technique, Materials Research Bulletin, 46 (2011) 643-648.

[103] C.-J. Li, B. Wang, J.-N. Wang, Magnetic and microwave absorbing properties of electrospun $Ba_{(1-x)}La_xFe_{12}O_{19}$ nanofibers, Journal of Magnetism and Magnetic Materials, 324 (2012) 1305-1311.

[104] Y.-M. Kang, High saturation magnetization in La–Ce–Zn–doped M-type Sr-hexaferrites, Ceramics International, 41 (2015) 4354-4359.

[105] L. Peng, L. Li, R. Wang, Y. Hu, X. Tu, X. Zhong, Microwave sintered $Sr_{1-x}La_xFe_{12-x}Co_xO_{19}$ ($x$ = 0–0.5) ferrites for use in low temperature co-fired ceramics technology, Journal of Alloys and Compounds, 656 (2016) 290-294.

[106] S. Ounnunkad, Improving magnetic properties of barium hexaferrites by La or Pr substitution, Solid State Communications, 138 (2006) 472-475.

[107] M. Awawdeh, I. Bsoul, S.H. Mahmood, Magnetic properties and Mössbauer spectroscopy on Ga, Al, and Cr substituted hexaferrites, Journal of Alloys and Compounds, 585 (2014) 465-473.

[108] S. Wang, J. Ding, Y. Shi, Y. Chen, High coercivity in mechanically alloyed $BaFe_{10}Al_2O_{19}$, Journal of magnetism and magnetic materials, 219 (2000) 206-212.

[109] I. Ali, M. Islam, M. Awan, M. Ahmad, Effects of Ga–Cr substitution on structural and magnetic properties of hexaferrite ($BaFe_{12}O_{19}$) synthesized by sol–gel auto-combustion route, Journal of Alloys and Compounds, 547 (2013) 118-125.

[110] Joonghoe Dho, E.K. Lee, N.H.H. J.Y. Park, Effects of the grain boundary on the coercivity of barium ferrite $BaFe_{12}O_{19}$, Journal of Magnetism and Magnetic Materials, 285 (2005) 164-168.

[111] J. Dahal, L. Wang, S. Mishra, V. Nguyen, J. Liu, Synthesis and magnetic properties of $SrFe_{12-x-y}Al_xCo_yO_{19}$ nanocomposites prepared via autocombustion technique, Journal of Alloys and Compounds, 595 (2014) 213-220.

[112] B. Rai, S. Mishra, V. Nguyen, J. Liu, Synthesis and characterization of high coercivity rare-earth ion doped $Sr_{0.9}RE_{0.1}Fe_{10}Al_2O_{19}$ (RE: Y, La, Ce, Pr, Nd, Sm, and Gd), Journal of Alloys and Compounds, 550 (2013) 198-203.

[113] P. Kazin, L. Trusov, D. Zaitsev, Y.D. Tretyakov, M. Jansen, Formation of submicron-sized $SrFe_{12-x}Al_xO_{19}$ with very high coercivity, Journal of Magnetism and Magnetic Materials, 320 (2008) 1068-1072.

[114] D. Chen, Y. Liu, Y. Li, K. Yang, H. Zhang, Microstructure and magnetic properties of Al-doped barium ferrite with sodium citrate as chelate agent, Journal of magnetism and magnetic Materials, 337 (2013) 65-69.

[115] A.A. Nourbakhsh, M. Noorbakhsh, M. Nourbakhsh, M. Shaygan, K.J. Mackenzie, The effect of nano sized $SrFe_{12}O_{19}$ additions on the magnetic properties of chromium-doped strontium-hexaferrite ceramics, Journal of Materials Science: Materials in Electronics, 22 (2011) 1297-1302.

[116] S. Ounnunkad, P. Winotai, Properties of Cr-substituted M-type barium ferrites prepared by nitrate–citrate gel-autocombustion process, Journal of Magnetism and Magnetic Materials, 301 (2006) 292-300.

[117] S. Katlakunta, S.S. Meena, S. Srinath, M. Bououdina, R. Sandhya, K. Praveena, Improved magnetic properties of $Cr^{3+}$ doped $SrFe_{12}O_{19}$ synthesized via microwave hydrothermal route, Materials Research Bulletin, 63 (2015) 58-66.

[118] V.P. Singh, G. Kumar, R. Kotnala, J. Shah, S. Sharma, K. Daya, K.M. Batoo, M. Singh, Remarkable magnetization with ultra-low loss $BaGd_xFe_{12-x}O_{19}$ nanohexaferrites for applications up to C-band, Journal of Magnetism and Magnetic Materials, 378 (2015) 478-484.

[119] R. Pawar, S. Desai, Q. Tamboli, S.E. Shirsath, S. Patange, $Ce^{3+}$ incorporated structural and magnetic properties of M type barium hexaferrites, Journal of Magnetism and Magnetic Materials, 378 (2015) 59-63.

[120] P. Long, H. Yue-Bin, G. Cheng, L. Le-Zhong, W. Rui, H. Yun, T. Xiao-Qiang, Preparation and magnetic properties of $SrFe_{12}O_{19}$ ferrites suitable for use in self-biased LTCC circulators, Chinese Physics Letters, 32 (2015) 017502.

[121] G. Bate, Magnetic recording materials since 1975, Journal of Magnetism and Magnetic materials, 100 (1991) 413-424.

[122] G. Bate, Recording materials, in: P. E, Wohlfarth (Ed.) Ferromagnetic materials, North-Holland Publishing Company, New York, 1980, pp. 381-508.

[123] D. Han, Z. Yang, H. Zeng, X. Zhou, A. Morrish, Cation site preference and magnetic properties of Co-Sn-substituted Ba ferrite particles, Journal of magnetism and magnetic materials, 137 (1994) 191-196.

[124] A. Gonzalez-Angeles, G. Mendoza-Suarez, A. Gruskova, I. Toth, V. Jančárik, M. Papanova, J. Escalante-Garcí, Magnetic studies of NiSn-substituted barium hexaferrites processed by attrition milling, Journal of magnetism and magnetic materials, 270 (2004) 77-83.

[125] A. González-Angeles, G. Mendoza-Suárez, A. Grusková, J. Sláma, J. Lipka, M. Papánová, Magnetic structure of $Sn^{2+}Ru^{4+}$-substituted barium hexaferrites prepared by mechanical alloying, Materials letters, 59 (2005) 1815-1819.

[126] A. González-Angeles, G. Mendoza-Suarez, A. Grusková, M. Papanova, J. Slama, Magnetic studies of Zn–Ti-substituted barium hexaferrites prepared by mechanical milling, Materials letters, 59 (2005) 26-31.

[127] A. González-Angeles, G. Mendoza-Suarez, A. Grusková, J. Lipka, M. Papanova, J. Slama, Effect of (Ni, Zn) Ru mixtures on magnetic properties of barium hexaferrites yielded by high-energy milling, Journal of magnetism and magnetic materials, 285 (2005) 450-455.

[128] I. Bsoul, S.H. Mahmood, A.F. Lehlooh, A. Al-Jamel, Structural and magnetic properties of $SrFe_{12-2x}Ti_xRu_xO_{19}$, Journal of Alloys and Compounds, 551 (2013) 490-495.

[129] G.H. Dushaq, S.H. Mahmood, I. Bsoul, H.K. Juwhari, B. Lahlouh, M.A. AlDamen, Effects of molybdenum concentration and valence state on the structural and magnetic properties of $BaFe_{11.6}Mo_xZn_{0.4-x}O_{19}$ hexaferrites, Acta Metallurgica Sinica (English Letters), 26 (2013) 509-516.

[130] S.H. Mahmood, G.H. Dushaq, I. Bsoul, M. Awawdeh, H.K. Juwhari, B.I. Lahlouh, M.A. AlDamen, Magnetic Properties and Hyperfine Interactions in M-Type $BaFe_{12-2x}Mo_xZn_xO_{19}$ Hexaferrites, Journal of Applied Mathematics and Physics, 2 (2014) 77-87.

[131] H. Vincent, E. Brando, B. Sugg, Cationic Distribution in Relation to the Magnetic Properties of New M-Hexaferrites with Planar Magnetic Anisotropy $BaFe_{12-2x}Ir_xMe_xO_{19}$ (Me= Co, Zn, x≈ 0.85 and x≈ 0.50), Journal of Solid State Chemistry, 120 (1995) 17-22.

[132] B. Sugg, H. Vincent, Magnetic properties of new M-type hexaferrites $BaFe_{12-2x}Ir_xCo_xO_{19}$, Journal of magnetism and magnetic materials, 139 (1995) 364-370.

[133] M.V. Rane, D. Bahadur, S. Mandal, M. Patni, Characterization of $BaFe_{12-2x}Co_xZr_xO_{19}$ ($0\leq x\leq 0.5$) synthesised by citrate gel precursor route, Journal of magnetism and magnetic materials, 153 (1996) L1-L4.

[134] S. Sugimoto, K. Okayama, S.-i. Kondo, H. Ota, M. Kimura, Y. Yoshida, H. Nakamura, D. Book, T. Kagotani, M. Homma, Barium M-type ferrite as an electromagnetic microwave absorber in the GHz range, Materials Transactions, JIM, 39 (1998) 1080-1083.

[135] H. Fang, Z. Yang, C. Ong, Y. Li, C. Wang, Preparation and magnetic properties of (Zn–Sn) substituted barium hexaferrite nanoparticles for magnetic recording, Journal of magnetism and magnetic materials, 187 (1998) 129-135.

[136] P. Wartewig, M. Krause, P. Esquinazi, S. Rösler, R. Sonntag, Magnetic properties of Zn- and Ti-substituted barium hexaferrite, Journal of magnetism and magnetic materials, 192 (1999) 83-99.

[137] F. Wei, H. Fang, C. Ong, C. Wang, Z. Yang, Magnetic properties of $BaFe_{12-2x}Zn_xZr_xO_{19}$ particles, Journal of Applied Physics, 87 (2000) 8636-8639.

[138] G. Mendoza-Suarez, L. Rivas-Vazquez, J. Corral-Huacuz, A. Fuentes, J. Escalante-Garcí, Magnetic properties and microstructure of $BaFe_{11.6-2x}Ti_xM_xO_{19}$ (M= Co, Zn, Sn) compounds, Physica B: Condensed Matter, 339 (2003) 110-118.

[139] A.M. Alsmadi, I. Bsoul, S.H. Mahmood, G. Alnawashi, K. Prokeš, K. Siemensmeyer, B. Klemke, H. Nakotte, Magnetic study of M-type doped barium hexaferrite nanocrystalline particles, Journal of Applied Physics, 114 (2013) 243910.

[140] Z. Yang, C. Wang, X. Li, H. Zeng, (Zn, Ni, Ti) substituted barium ferrite particles with improved temperature coefficient of coercivity, Materials Science and Engineering: B, 90 (2002) 142-145.

[141] S. Pignard, H. Vincent, E. Flavin, F. Boust, Magnetic and electromagnetic properties of RuZn and RuCo substituted $BaFe_{12}O_{19}$, Journal of magnetism and magnetic materials, 260 (2003) 437-446.

[142] S. Nilpairach, W. Udomkichdaecha, I. Tang, Coercivity of the co-precipitated prepared hexaferrites, $BaFe_{12-2x}Co_xSn_xO_{19}$, Journal of the Korean Physical Society, 48 (2006) 939-945.

[143] O. Kubo, E. Ogawa, Barium ferrite particles for high density magnetic recording, Journal of magnetism and magnetic materials, 134 (1994) 376-381.

[144] X. Batlle, X. Obradors, J. Rodriguez-Carvajal, M. Pernet, M. Cabanas, M. Vallet, Cation distribution and intrinsic magnetic properties of Co-Ti-doped M-type barium ferrite, Journal of applied physics, 70 (1991) 1614-1623.

[145] X. Zhou, A. Morrish, Z. Li, Y. Hong, Site preference for $Co^{2+}$ and $Ti^{4+}$ in Co-Ti substituted barium ferrite, Magnetics, IEEE Transactions on, 27 (1991) 4654-4656.

[146] A. Gruskova, J. Slama, M. Michalikova, J. Lipka, I. Toth, P. Kaboš, Preparation of substituted barium ferrite powders, Journal of magnetism and magnetic materials, 101 (1991) 227-229.

[147] A. Morrish, X. Zhou, Z. Yang, H.-X. Zeng, Substituted barium ferrites; sources of anisotropy, Hyperfine Interactions, 90 (1994) 365-369.

[148] Z. Šimša, S. Lego, R. Gerber, E. Pollert, Cation distribution in Co-Ti-substituted barium hexaferrites: a consistent model, Journal of magnetism and magnetic materials, 140 (1995) 2103-2104.

[149] G. Bottoni, Magnetization stability and interactions in particulate recording media, Materials chemistry and physics, 42 (1995) 45-50.

[150] K. Kakizaki, N. Hiratsuka, T. Namikawa, Fine structure of acicular $BaCo_xTi_xFe_{12-2x}O_{19}$ particles and their magnetic properties, Journal of magnetism and magnetic materials, 176 (1997) 36-40.

[151] Y. Li, R. Liu, Z. Zhang, C. Xiong, Synthesis and characterization of nanocrystalline $BaFe_{9.6}Co_{0.8}Ti_{0.8}M_{0.8}O_{19}$ particles, Materials chemistry and physics, 64 (2000) 256-259.

[152] G. Mendoza-Suarez, J. Corral-Huacuz, M. Contreras-García, H. Juarez-Medina, Magnetic properties of $BaFe_{11.6-2x}Co_xTi_xO_{19}$ particles produced by sol–gel and spray-drying, Journal of magnetism and magnetic materials, 234 (2001) 73-79.

[153] M. Kuznetsov, Q. Pankhurst, I. Parkin, Novel SHS routes to CoTi-doped M-type ferrites, Journal of Materials Science: Materials in Electronics, 12 (2001) 533-536.

[154] A. Gruskova, J. Slama, R. Dosoudil, D. Kevicka, V. Jančárik, I. Toth, Influence of Co–Ti substitution on coercivity in Ba ferrites, Journal of magnetism and Magnetic Materials, 242 (2002) 423-425.

[155] S.Y. An, I.-B. Shim, C.S. Kim, Mössbauer and magnetic properties of Co–Ti substituted barium hexaferrite nanoparticles, Journal of applied physics, 91 (2002) 8465-8467.

[156] C. Wang, L. Li, J. Zhou, X. Qi, Z. Yue, High-frequency magnetic properties of Co-Ti substituted barium ferrites prepared by modified chemical coprecipitation method, Journal of Materials Science: Materials in Electronics, 13 (2002) 713-716.

[157] C. Wang, X. Qi, L. Li, J. Zhou, X. Wang, Z. Yue, High-frequency magnetic properties of low-temperature sintered Co-Ti substituted barium ferrites, Materials Science and Engineering: B, 99 (2003) 270-273.

[158] Z. Haijun, L. Zhichao, M. Chenliang, Y. Xi, Z. Liangying, W. Mingzhong, Preparation and microwave properties of Co-and Ti-doped barium ferrite by citrate sol–gel process, Materials chemistry and physics, 80 (2003) 129-134.

[159] R. Lima, M.S. Pinho, M.L. Gregori, R.R. Nunes, T. Ogasawara, Effect of double substituted *m*-barium hexaferrites on microwave absorption properties, Materials Science-Poland, 22 (2004) 245-252.

[160] J. Rodriguez-Carvajal, FULLPROF 98. Program for Rietveld Pattern Matching Analysis of Powder Patterns, unspublished results, Grenoble, 1998.(b) Rodriguez-Carvajal, J, Physica B, 55 (1993) 192.

[161] X. Obradors, X. Solans, A. Collomb, D. Samaras, J. Rodriguez, M. Pernet, M. Font-Altaba, Crystal structure of strontium hexaferrite $SrFe_{12}O_{19}$, Journal of Solid State Chemistry, 72 (1988) 218-224.

[162] O. Kalogirou, G. Haack, B. Röhl, W. Gunβer, Mössbauer study of a modified M-type Ba (Sr)-ferrite prepared by ion exchange, Solid state ionics, 63 (1993) 528-533.

[163] G.B. Teh, Y.C. Wong, R.D. Tilley, Effect of annealing temperature on the structural, photoluminescence and magnetic properties of sol–gel derived Magnetoplumbite-type

(M-type) hexagonal strontium ferrite, Journal of Magnetism and Magnetic Materials, 323 (2011) 2318-2322.

[164] B.E. Warren, X-ray Diffraction, Addison-Wesley, Reading, Massachsetts, 1969.

[165] S. Masoudpanah, S.S. Ebrahimi, Fe/Sr ratio and calcination temperature effects on processing of nanostructured strontium hexaferrite thin films by a sol–gel method, Research on Chemical Intermediates, 37 (2011) 259-266.

[166] B.D. Cullity, C.D. Graham, Introduction to magnetic materials, John Wiley & Sons2011.

[167] S.H. Mahmood, A. Awadallah, Y. Maswadeh, I. Bsoul, Structural and magnetic properties of Cu-V substituted M-type barium hexaferrites, IOP Conference Series: Materials Science and Engineering, IOP Publishing, 2015, pp. 012008.

[168] Q. Pankhurst, Anisotropy field measurement in barium ferrite powders by applied field Mossbauer spectroscopy, Journal of Physics: Condensed Matter, 3 (1991) 1323.

[169] H. Pfeiffer, W. Schüppel, Investigation of magnetic properties of barium ferrite powders by remanence curves, physica status solidi (a), 119 (1990) 259-269.

[170] S.H. Mahmood, I. Bsoul, Hopkinson peak and superparamagnetic effects in $BaFe_{12-x}Ga_xO_{19}$ nanoparticles, EPJ Web of Conferences, 29 (2012) 00039.

[171] P. Kelly, K. O'Grady, P. Mayo, R. Chantrell, Switching mechanisms in cobalt-phosphorus thin films, Magnetics, IEEE Transactions on, 25 (1989) 3881-3883.